\documentclass[aps,prd,nofootinbib]{revtex4}
\usepackage[utf8]{inputenc}
\usepackage[english]{babel}
\usepackage[T1]{fontenc}

\usepackage{amsmath}
\usepackage{amsfonts}
\usepackage{amssymb}
\usepackage{graphicx}

\usepackage{lmodern}

\usepackage{hyperref}
\hypersetup{colorlinks, 
citecolor=blue,
linkcolor=red,
urlcolor=black}

\usepackage[normalem]{ulem}

\usepackage[title]{appendix}
\usepackage{braket}
\usepackage[center]{caption} 
\usepackage{subfigure}

\begin{document}

\title{Bosonic Casimir effect in an aether-like Lorentz-violating scenario with higher order derivatives}
\author{R. A. Dantas}
\email{robson.dantas@academico.ufpb.br}
\author{H. F. Santana Mota}
\email{hmota@fisica.ufpb.br}
\author{E. R. Bezerra de Mello}
\email{emello@fisica.ufpb.br}
\affiliation{Departamento de F\'isica, Universidade Federal da Para\'iba, Caixa Postal 5008, Jo\~ao Pessoa, Para\'iba, Brazil}

\begin{abstract}
In this paper we investigate the bosonic Casimir effect in a Lorentz-violating symmetry scenario. The theoretical model adopted consists of a real massive scalar quantum field confined in a region between two large parallel plates, having its dynamics governed by a modified Klein-Gordon equation that presents a Lorentz symmetry breaking term. In this context  we admit that the quantum field obeys specific boundary conditions on the plates. The Lorentz-violating symmetry is implemented 
by the presence of an arbitrary constant space-like vector in a CPT-even aether-like approach, considering a direct coupling between this vector with the derivative of the field in higher order. The modification on the Klein-Gordon equation produces important corrections on the Casimir energy and pressure. Thus, we show that these corrections strongly depend on the order of the higher derivative term and the specific direction of the constant vector, as well as the boundary conditions considered.
\end{abstract}

\maketitle

\bigskip


\bigskip


\bigskip

\newpage

\section{Introduction}%%%%%%%%%%%%%%%%%%%%%%%%%%%%%%%%%%%%%%%%%%%%%%%%%%%%%%%%%%%%%%%%%%%%%%%%%%%%%%%%

 The Casimir effect is one of the most important macroscopic consequences of the existence of the quantum vacuum. Although being theoretically proposed in 1948 by H. B. Casimir \cite{Casimir_48}, the theoretical result was only verified to be compatible with experiments $10$ years later by M. J. Sparnaay \cite{Sparnaay_58}. In the 90's, experiments have confirmed the Casimir effect with high degree of accuracy \cite{Lamoureux_97,Mohideen_98}.  In his original work, Casimir predicted that due to the quantum vacuum  fluctuations associated with the electromagnetic field, two parallel flat neutral (grounded) plates, separated by a distance $a$, attract each other with a force per unit area given by:
 \begin{equation}
 	\frac FA = - \frac{\pi^2 \hbar c}{240 a^4}\   ,
 \end{equation}
 where $A$ is the area of the plates. 
 
 In general, the Casimir effect is defined as being a force per unit area, when boundary conditions are imposed 
 on quantum fields. The simplest theoretical device to study the Casimir effect is constituted by two neutral parallel plates placed in the (classical) vacuum. As the quantum vacuum consists of an infinite set of waves that contemplates all possible wavelengths, when the plates are considered, only a few wavelengths are allowed between them.
 
 The Lorentz invariance, which is as a cornerstone of Quantum Field Theory, has been questioned in a work by V. A. Kostelecky and S. Samuel \cite{Kostelecky_88}, that describes a mechanism in string theory that allows the violation of Lorentz symmetry at the Planck energy scale. According to this mechanism, the violation of the Lorentz symmetry is introduced by the emergence of non-vanishing vacuum expectation values of some vector and tensor components, which imply preferential directions, providing in this way a space-time anisotropy. In the quantum gravity context, Ho\u rava-Lifshitz (HL) proposed a theory \cite{Horava_09} with the objective to implement  possible convergence of quantum corrections, by imposing different properties of scales in which coordinates space and time are set. The HL approach clearly provides an anisotropy between space and time, and can be applied not only to gravity, but also to other field theory models, including scalar, spinor and gauge theories.  Among the most important results achieved in these scenarios, one can emphasize the calculation of the one-loop  effective potential in HL-like QED and HL-like Yukawa model \cite{EP} and the study of different issues related to the renormalization of these theories \cite{ren}. If there is a violation of the Lorentz symmetry at the Planck energy scale in a more fundamental theory, the effects of this breakdown must manifest itself in other energy scales in different QFT models. Other mechanisms of violation of Lorentz symmetry are possible, such as space-time non-commutativity \cite{Carroll:2001ws, Anisimov:2001zc, Carlson:2001sw, Hewett:2000zp, Bertolami:2003nm}.

The violation of the Lorentz symmetry became of great experimental interest. For instance, the high accuracy experimental measurements of the Casimir pressure became a great allied in the study of the Lorentz symmetry breakdown in theoretical models within Field Theory, in the search of vestiges left by the violation. 

The first analysis of the Casimir energy in the Lorentz-violating (LV) theories have been developed  in \cite{FrankTuran,Escobar} considering different Lorentz-breaking extensions of the QED. In addition, the studies of Casimir effects associated with massless scalar and fermionic quantum fields confined in the region between two large parallel plates taken into account the HL formalism, has been investigated in Refs. \cite{Ulion:2015kjx} and \cite{Deivid}, respectively. More recently the Casimir effect associated with a massive real scalar field was developed in Ref. \cite{Maluf2020}. 

Considering direct coupling between the derivative of the field with an arbitrary constant four-vector in an aether-like CPT-even Lorentz symmetry breaking, the analysis of Casimir effects associated with real scalar and fermionic massive fields, has been investigated in Refs. \cite{Messias2017} and \cite{Messias2019}, respectively. Moreover, local Casimir densities in a LV scenario have been analyzed in \cite{Escobar1,Escobar2}. In Ref. \cite{Erdas2020} it was considered the influence of a constant magnetic field on the Casimir effect in the Lorentz violating scalar field. The thermal effect on the Casimir energy and pressure caused by the Lorentz violating scalar field, was investigated in \cite{Messias2018}.
The analysis of the Casimir energy and topological mass associated with a massive scalar field in LV scenario, were considered in Refs. \cite{Messias2020, HJ}. 

In this paper we intend to continue in the same line of investigation, i.e., analyzing the Casimir effect associated with a massive scalar quantum field in a LV scenario; however at this time we shall take into consideration that the LV is implemented by a new term that involves higher order derivatives of the field, coupled to a space-like constant vector. In this way we may understand this term as a combination of the HL methodology with the aether-like CPT-even Lorentz symmetry breaking. 

This paper is organized as follows: In Section \ref{Sect2}, we briefly introduce the theoretical model that governs the dynamics of the real scalar field. We present the LV bosonic action and the corresponding modified Klein-Gordon equation. In Section \ref{Sect3} we develop the  calculation of the Casimir energies, in cases where the constant vector is parallel and orthogonal to the plates. In order to confine the bosonic field between the two parallel plates, we should impose that the flux of virtual particles crossing the plates is zero. This can be done by imposing  Dirichlet, Neumann or Mixed boundary conditions on the field at the plates. Finally we leave for Conclusions \ref{Concl} our most relevant remarks found in this paper. Here, units are assumed to be $\hbar=c=1$, and the metric 
signature will be taken as $(+,-,-,-)$.

\section{Klein-Gordon equation in aether-like Lorentz symmetry violation scenario with higher order derivatives}%%%%%%%%%%%%%%%%%%%%%%%%%%%%%%%%%%%%%%%
\label{Sect2}
In this section we introduce the theoretical model that we want to investigate. It is composed by a massive scalar
 quantum field in a Lorentz-violating symmetry scenario introduced by the presence of a constant space-like vector, in a aether-like approach, and considering its direct coupling with the derivative of the field in higher order. In this sense the Lorentz violation symmetry is caused by the presence of a constant background vector, and by an anisotropy between space and time coordinates by a scaling transformations. This model is formally given by the Lagrangian density below:
\begin{equation}
\label{L}
\mathcal{L} = \frac{1}{2} \left[ (\partial_{\mu}\phi)(\partial^{\mu}\phi)
 - l^{2(\epsilon-1)}(-1)^{\epsilon}[(u^\mu\partial_\mu)^{\epsilon}\phi]^2-m^{2}\phi^{2} \right] \  .
\end{equation}
In the above Lagrangian, the parameter $l$ is of order of the inverse of the energy scale where the Lorentz symmetry is broken. The dimensionless constant vector $u^\mu$, that is associated with a preferential direction, couples to the scalar field through its derivative as explained above, and the parameter $\epsilon$ is an integer number. 

In this formalism the  modified Klein-Gordon equation (KG) reads,
\begin{equation}
\label{kgmodific}
\left [\square+l^{2(\epsilon-1)}(u^\mu\partial_\mu)^{2\epsilon}+m^{2}\right]\phi=0  \   .
\end{equation}
For $\epsilon=1$ the above equation coincides with the one presented in \cite{Messias2017}. In the latter, the analysis of the Casimir energy and pressure have been considered admitting that the constant four vector is both time-like and space-like, separately. Because we are interested in investigating the behavior of the scalar field in higher order derivatives theory, we will consider $\epsilon\geq2$. Moreover, to avoid unitarity problem we will also assume that the vector $u^\mu$ is only space-like. Another way to implement  higher-order derivative in a LV scenario is by considering higher-order time derivative of the field; however the presence of this term may violate unitarity of the theory.

Imposing that the action associated with the Lagrangian \eqref{L} is invariant under the infinitesimal translation, $x^\mu\to x^\mu+\delta a^\mu$,  the obtained energy-momentum tensor (EM) reads,\footnote{In fact  our derivation of EM  was developed by an induction procedure, i.e, we first assumed $\epsilon=2$, followed by $\epsilon=3$ and $\epsilon=4$.}
\begin{eqnarray}
	\label{Lagran}
T^{\mu\nu}&=&\frac{\partial \mathcal{L}}{\partial(\partial_{\mu }\phi)}\partial^{\nu }\phi + \frac{\partial \mathcal{L}}{\partial(\partial_{\mu_{1}}...\partial_{\mu_{\epsilon -1}}\partial _{\mu }\phi)}\partial_{\mu_{1}}...\partial_{\mu_{\epsilon -1}}\partial^{\nu }\phi - \partial _{\mu _{1}}\frac{\partial \mathcal{L}}{\partial(\partial_{\mu_{1}}...\partial_{\mu_{\epsilon -1}}\partial _{\mu }\phi)}\partial_{\mu_{2}}...\partial_{\mu_{\epsilon -1}}\partial^{\nu }\phi \nonumber\\&+& \partial _{\mu _{1}}\partial _{\mu _{2}}\frac{\partial \mathcal{L}}{\partial(\partial_{\mu_{1}}...\partial_{\mu_{\epsilon -1}}\partial _{\mu }\phi)}\partial_{\mu_{3}}...\partial_{\mu_{\epsilon -1}}\partial^{\nu }\phi + ... + (-1)^{\epsilon -1}\partial _{\mu _{1}}...\partial _{\mu _{\epsilon -1}}\frac{\partial \mathcal{L}}{\partial(\partial_{\mu_{1}}...\partial_{\mu_{\epsilon -1}}\partial _{\mu }\phi)}\partial^{\nu }\phi - \eta ^{\mu \nu }\mathcal{L}  \  .
\end{eqnarray}

Substituting the Lagrangian \eqref{L} into \eqref{Lagran}, we obtain: 
\begin{eqnarray}
	\label{Lagran_1}
T^{\mu\nu} &=& (\partial ^{\mu }\phi) (\partial ^{\nu }\phi)  + \epsilon !l^{2(\epsilon-1)}u^{\mu }\Bigg\{ \left [ (u\cdot \partial )^{2\epsilon -1}\phi  \right ](\partial ^{\nu }\phi)  - \left [ (u\cdot \partial )^{2\epsilon -2}\phi  \right ](u\cdot \partial )(\partial ^{\nu }\phi) \nonumber\\&+& \left [ (u\cdot \partial )^{2\epsilon -3}\phi  \right ](u\cdot \partial )^{2}(\partial ^{\nu }\phi) +...- (-1)^{\epsilon }\left [ (u\cdot \partial )^{\epsilon }\phi  \right ](u\cdot \partial )^{\epsilon -1}(\partial ^{\nu }\phi) \Bigg\} - \eta ^{\mu \nu }\mathcal{L}  \  .
\end{eqnarray}
Although by construction, the general expression for the energy-momentum tensor, Eq. \eqref{Lagran}, satisfies the condition, 
\begin{eqnarray}
	\label{Cons}
	\partial_\mu T^{\mu\nu} = 0 \ ,
\end{eqnarray}	
we have explicitly shown, for different values of $\epsilon$,  that the above equation is obeyed by the Lagrangian \eqref{L}.

Furthermore, we can see that  the energy-momentum  tensor is not symmetric: its anti-symmetric part is given by
\begin{eqnarray}
T^{\mu \nu }-T^{\nu \mu } &=& \epsilon !l^{2(\epsilon-1)}\Bigg\{ \left [ (u\cdot \partial )^{2\epsilon -1}\phi  \right ]  - \left [ (u\cdot \partial )^{2\epsilon -2}\phi  \right ](u\cdot \partial ) \nonumber\\&+& \left [ (u\cdot \partial )^{2\epsilon -3}\phi  \right ](u\cdot \partial )^{2} +...- (-1)^{\epsilon }\left [ (u\cdot \partial )^{\epsilon }\phi  \right ](u\cdot \partial )^{\epsilon -1} \Bigg\} \left [ u^{\mu }(\partial ^{\nu }\phi ) - u^{\nu  }(\partial ^{\mu }\phi )\right ] \  .
\end{eqnarray}
This anti-symmetry is typical for Lorentz-symmetry violation formalism. Moreover, we have explicitly checked that $\partial_\nu T^{\mu\nu}\neq 0$. 

\section{The Casimir effect in the context of higher order derivatives Lorentz symmetry violation}
\label{Sect3}

\begin{figure}[h]
\centering
\includegraphics[height=5cm, width=8cm]{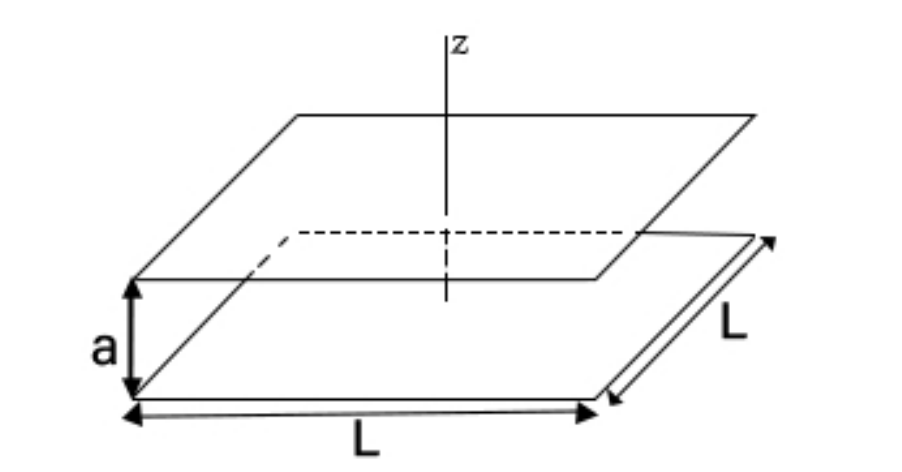}
\caption{Two parallel plates with area $L^2$ separated by a distance $a<<L$.}
\label{fig1}
\end{figure}

The main objective of this section is to analyze how the LV symmetry represented by the presence of a higher order derivative term of the field, along a specific direction, modifies the dispersion relations responsible from the deviation of the Casimir energy and pressure, when compared with the scenario preserving the Lorentz symmetry. As we have already mentioned we will assume that the scalar field operator, $\hat{\phi}(x)$, satisfies specific boundary conditions on the plates exhibited in Fig. \ref{fig1}.  

In order to obtain the Hamiltonian operator, $\hat{H}$, we have first to calculate the normalized set of positive/negative energy solutions of Eq. \eqref{kgmodific} that obey specific boundary conditions on the plates. Doing this we can calculate the total vacuum energy of the system and then determine the Casimir energy for each case considered. 

\subsection{Dirichlet condition}%%%%%%%%%%%%%%%%%%%%%%%%%%%%%%%%%%%%%%%%%%%%%%%%%%%%%%%%%%%%%%%%%%%%%%%%%%%
\label{Dirichlet}
It has been shown in \cite{Messias2017} that the quantum field operator, $ \hat{\phi}(x)$,  expressed in terms of the normalized positive/negative energy solution of the Klein-Gordon equation that satifies Dirichlet boundary condition on the plates, $z=0$ and $z=a$, that is,
\begin{equation}
\phi(x)_{z=0} =\phi(x)_{z=a} = 0 \  ,
\end{equation}
 has the general form given below,
\begin{eqnarray}
	\label{Field_Diri}
\hat{\phi}(x) = \int \mathrm{d}^2\textbf{k} \sum_{n=1}^{\infty} 
\frac{1}{[(2\pi)^2 a\ \omega_{\textbf{k},n}]^{1/2}} \sin\left(\dfrac{n\pi}
{a}z\right)[\hat{a}_{\textbf{k},n}e^{-i k x} + \hat{a}^{\dagger}_{\textbf{k},n}e^{i k x}] \  ,
\end{eqnarray}
where $\hat{a}_{\textbf{k},n}$ and $\hat{a}^{\dagger}_{\textbf{k},n}$ correspond to the 
annihilation and creation operators, respectively. These operators satisfy the following commutation relations 
\begin{equation}
\label{algebra}
\begin{split}
[\hat{a}_{\textbf{k},n}, \hat{a}^{\dagger}_{\textbf{k}',n'}] & = \delta_{n,n'}\delta^2(\textbf{k}-\textbf{k}') , \\
[\hat{a}_{\textbf{k},n}, \hat{a}_{\textbf{k}',n'}] = & [\hat{a}^{\dagger}_{\textbf{k},n}, \hat{a}^{\dagger}_{\textbf{k}',n'}] = 0 \  ,
\end{split}
\end{equation}
being $kx \equiv \omega_{\textbf{k},n}t - k_{x}x - k_{y}y - k_zz$. The explicit form of $\omega_{\textbf{k},n}$ will depend on the specific model adopted for the LV term. This subject will be explored in the following subsections.

\subsubsection{Vector parallel to the plates}%%%%%%%%%%%%%%%%%%%%%%%%%%%%%%%%%%%%%%%%%%%%%%%%%%%%%%%%%%%%%%%%%%%%
The space-like four-vector, $u^{\mu}$, can be in three different directions, parallel to the plates, $u^{\mu}=(0,1,0,0)$ and $u^{\mu}=(0,0,1,0)$, and  perpendicular to the plates, $u^{\mu}=(0,0,0,1)$. The dispersion relations associated with the first two vectors are the same. Thereby, here in this subsection, we will consider a parallel vector as being, 
\begin{eqnarray}
u^{\mu}=(0,1,0,0) \  .
\end{eqnarray}
The corresponding dispersion relation is
\begin{eqnarray}
\omega_{\textbf{k},n}^2 = k_{x}^2 + k_{y}^2 + l^{2(\epsilon-1)}(-1)^{\epsilon}k_{x}^{2\epsilon} + \left(\frac{n\pi}{a}\right)^2 + m^2 \ .
\end{eqnarray}
Hence, the Hamiltonian operator $\hat{H}$, resulting from the canonical quantization reads
\begin{eqnarray}
	\label{Hamilt}
\hat{H} = \frac{1}{2}\int\mathrm{d}^2\textbf{k} \sum_{n=1}^{\infty}\omega_{\textbf{k},n}\left[2 \hat{a}^{\dagger}_{\textbf{k,n}} \hat{a}_{\textbf{k,n}} + 
\frac{L^2}{(2\pi)^2}\right] \   .
\end{eqnarray}

The vacuum energy is obtained by taking the vacuum expectation value of $\hat{H}$:
\begin{eqnarray}
\label{vacuumdpa}
E_{0} = \bra{0}\hat{H}\ket{0} = \frac{L^2}{8 \pi^2}\int \mathrm{d}^2\textbf{k} 
\sum_{n=1}^{\infty}\omega_{\textbf{k},n} \  .
\end{eqnarray}
Performing a change of coordinates from Cartesian coordinate $(k_{x},k_{y})$ to polar one, $(k,\theta)$, and making a change of variable $u=ak$, we obtain
\begin{eqnarray}
	\label{Casimir_E_1}
E_{0}=\frac{L^2}{8 \pi^2a^{3}}\int_{0}^{2\pi}d\theta\int_{0}^{\infty}udu
\sum_{n=1}^{\infty}\left [ u^{2}+(n\pi)^{2}+(ma)^{2}+\left ( \frac{l}{a} \right )^{2(\epsilon-1)}(-1)^{\epsilon}u^{2\epsilon}\cos^{2\epsilon}{\theta}\right ]^{\frac{1}{2}} \  .
\end{eqnarray}

Because it is our interest to investigate the LV correction on the Casimir energy due higher-order space derivative term, we will consider $\epsilon\ge2$. Although for this case the integral over the variable $u$ can be evaluated, the result is not very enlightening. Also, we have not found in literature the integral over $\theta$ for general values of $\epsilon$, even for $\epsilon=2$.  So, in order to provide some quantitative result for the correction on the Casimir energy caused by the Lorentz violating term, we develop an expansion in the parameter associated with the Lorentz violation. By doing an expansion up to the first order in the parameter $\frac{l}{a}\ll 1$, the expression \eqref{Casimir_E_1} can be written as
\begin{eqnarray}
 E_{0}&\approx&\frac{L^2}{8\pi^2a^{3}}\int_{0}^{2\pi}d\theta\int_{0}^{\infty}udu\sum_{n=1}^{\infty}\Bigg\{\left[u^{2}+(n\pi)^{2}+(ma)^{2}\right ]^{\frac{1}{2}}\nonumber\\ 
&+&\frac{1}{2}\left(\frac{l}{a}\right)^{2(\epsilon-1)}(-1)^{\epsilon}u^{2\epsilon}\cos^{2\epsilon}{\theta}\left [ u^{2}+(n\pi)^{2}+(ma)^{2}\right ]^{-\frac{1}{2}}\Bigg\}\ ,
\end{eqnarray}
where the first term is associated with the vacuum energy without Lorentz violation. Thus, after integration over the angular coordinate, the LV term becomes
\begin{eqnarray}
\label{Casimir_E_LV_1}
\tilde{E_{0}}=\frac{L^2}{8 \pi a^{3}}\left ( \frac{l}{a} \right )^{2(\epsilon-1)}\frac{(-1)^{\epsilon}(2\epsilon-1)!!}{(2\epsilon)!!}\int_{0}^{\infty}u^{^{(2\epsilon+1)}}\mathrm{d}u
\sum_{n=1}^{\infty}\left[u^{2}+(n\pi)^{2}+(ma)^{2}\right ]^{-\frac{1}{2}} \  .
\end{eqnarray}

The Casimir energy by unit area associated with a massive scalar quantum field confined between two large and parallel plates of area $L^2$, separated by a distance $a$, that obeys the Dirichlet boundary condition, has been obtained in \cite{Messias2017} in an integral representation by,
\begin{eqnarray}
	\label{int1}
	\frac{E_{C}}{L^{2}}=-\frac{am^4}{6\pi^{2}}\int_{1}^{\infty}\frac{(v^{2}-1)^{\frac{3}{2}}dv}
	{e^{2amv}-1}  \  .
\end{eqnarray}

Because our main interest in this research is to investigate the contribution of the Lorentz symmetry breaking in the Casimir energy, we will focus our analysis on Eq. \eqref{Casimir_E_LV_1}. As our first step to evaluate this contribution we will use the Abel-Plana summation formula below \cite{Bordag:2009zzd}, to develop the summation over the quantum number $n$. i.e.,
\begin{eqnarray}
	\label{Abel}
	\sum_{n=0}^{\infty}F(n) = \frac{1}{2}F(0) + \int_{0}^{\infty} F(t)\mathrm{d}t + 
	i \int_{0}^{\infty}\frac{\mathrm{d}t}{e^{2\pi t}-1}[ F(it) - F(-i t)] \  .
\end{eqnarray}

Then the expression \eqref{Casimir_E_LV_1} becomes
\begin{eqnarray}
\label{energywhithabeldpa}
\tilde{E_{0}}&=&\frac{L^2}{8 \pi a^{3}}\left ( \frac{l}{a} \right )^{2(\epsilon-1)}\frac{(-1)^{\epsilon}(2\epsilon-1)!!}{(2\epsilon)!!}\int_{0}^{\infty}u^{^{(2\epsilon+1)}}\mathrm{d}u\Bigg\{-\frac{1}{2}F(0)+\int_{0}^{\infty} 
F(t)\mathrm{d}t\nonumber\\&+&i\int_{0}^{\infty}\frac{F(it)-F(-it)}{e^{2\pi t}-1}\mathrm{d}t\Bigg\}\   ,
\end{eqnarray}
where 
\begin{eqnarray}
F(n) = \left[u^{2}+(n\pi)^{2}+(ma)^{2}\right ]^{-\frac{1}{2}} \  . 
\end{eqnarray}
Note that the first term on the right-hand side of \eqref{energywhithabeldpa} refers to the vacuum energy in the presence of only one plate, while the second term refers to the vacuum energy without boundary. Both  terms are divergent and do not contribute to the Casimir energy. As a result, the LV contribution to the Casimir energy per unit area of the plates is given by
\begin{eqnarray}
\frac{\tilde{E_{C}}}{L^2} &=& \frac{i}{8 \pi a^{3}}\left ( \frac{l}{a} \right )^{2(\epsilon-1)}\frac{(-1)^{\epsilon}(2\epsilon-1)!!}{(2\epsilon)!!}\int_{0}^{\infty}u^{^{(2\epsilon+1)}}\mathrm{d}u\nonumber\\&\times&\int_{0}^{\infty}\mathrm{d}t \frac{[u^2 + 
(it\pi)^{2}+(ma)^{2}]^{-1/2} -[u^2 + 
(-it\pi)^{2}+(ma)^{2}]^{-1/2}}{e^{2 \pi t} - 1} \  .
\end{eqnarray}
Performing a change of variable, with $t\pi = v$, we get
\begin{eqnarray}
	\label{E_C_3a}
\frac{\tilde{E_{C}}}{L^2} &=& \frac{i}{8 \pi^{2} a^{3}}\left ( \frac{l}{a} \right )^{2(\epsilon-1)}\frac{(-1)^{\epsilon}(2\epsilon-1)!!}{(2\epsilon)!!}\int_{0}^{\infty}u^{^{(2\epsilon+1)}}\mathrm{d}u\nonumber\\&\times&\int_{0}^{\infty}\mathrm{d}v\frac{[u^2 +(ma)^{2}+(iv)^2]^{-1/2} - 
[u^2 +(ma)^{2}+(-iv)^2]^{-1/2}}{e^{2v} - 1} \ . 
\end{eqnarray}

The integral over the variable $v$ must be considered in two cases, for $\left [ u^2 +(ma)^{2} \right ]^{1/2}>v$ and $\left [ u^2 +(ma)^{2} \right ]^{1/2}<v$. Taking into account that we have
\begin{itemize}
\item{For the case $\left [ u^2 +(ma)^{2} \right ]^{1/2}>v$}:
\end{itemize}
\begin{eqnarray}
[u^2 +(ma)^{2}+(\pm iv)^2]^{-1/2} = [u^2 +(ma)^{2}-v^2]^{-1/2}.
\end{eqnarray}
\begin{itemize}
\item{For the case $\left [ u^2 +(ma)^{2} \right ]^{1/2}<v$}:
\end{itemize}
 \begin{eqnarray}
[u^2 +(ma)^{2}+(\pm iv)^2]^{-1/2} = \mp i\left [v^{2}-(u^{2}+(ma)^{2}) \right ]^{-\frac{1}{2}}.
\end{eqnarray}
Consequently the integral in $u$ over the interval $[0,(u^2 +(ma)^{2})^{1/2}]$ vanishes. So, it remains
\begin{eqnarray}
	\label{E_C_3b}
\frac{\tilde{E_{C}}}{L^2} &=& \frac{1}{4\pi^{2} a^{3}}\left ( \frac{l}{a} \right )^{2(\epsilon-1)}\frac{(-1)^{\epsilon}(2\epsilon-1)!!}{(2\epsilon)!!}\int_{0}^{\infty}u^{^{(2\epsilon+1)}}\mathrm{d}u\nonumber\\&\times&\int_{\left [ u^{2}+(ma)^{2} \right ]^{\frac{1}{2}}}^{\infty}\mathrm{d}v\frac{[v^2-(u^{2}+(ma)^{2})]^{-1/2}}{e^{2v} - 1} \ .
\end{eqnarray}
Furthermore, performing the new change of variable $\rho^{2} =v^{2}-(u^{2}+(ma)^{2})$ , we find
\begin{eqnarray}
\frac{\tilde{E_{C}}}{L^2} &=& \frac{1}{4\pi^{2} a^{3}}\left ( \frac{l}{a} \right )^{2(\epsilon-1)}\frac{(-1)^{\epsilon}(2\epsilon-1)!!}{(2\epsilon)!!}\int_{0}^{\infty}u^{^{(2\epsilon+1)}}\mathrm{d}u\nonumber\\&\times&\int_{0}^{\infty}\frac{\mathrm{d}\rho }{\left [ \rho^{2}+u^{2}+(ma)^{2} \right ]^{\frac{1}{2}}\left [e^{2( \rho^{2}+u^{2}+(ma)^{2})^{\frac{1}{2}}} - 1\right ]} \ .
\end{eqnarray}
Now, by making a change of coordinates from the plane $(u,\rho)$ to the polar one, we get
\begin{eqnarray}
\label{energywithmassdpa}
\frac{\tilde{E_{C}}}{L^2} = \frac{1}{4\pi^{2} a^{3}}\left ( \frac{l}{a} \right )^{2(\epsilon-1)}\frac{(-1)^{\epsilon}}{(2\epsilon +1)}\int_{0}^{\infty}\frac{\sigma^{2(\epsilon+1)}\mathrm{d}\sigma}{\left [ \sigma^{2}+(ma)^{2} \right ]^{\frac{1}{2}}\left [e^{2(\sigma^{2}+(ma)^{2})^{\frac{1}{2}}} - 1\right ]} \ .
\end{eqnarray}

For massless field the above  integral  becomes,
\begin{eqnarray}
\frac{\tilde{E_{C}}}{L^2} = \frac{1}{4\pi^{2} a^{3}}\left ( \frac{l}{a} \right )^{2(\epsilon-1)}\frac{(-1)^{\epsilon}}{(2\epsilon +1)}\int_{0}^{\infty}\frac{\sigma^{2\epsilon+1}\mathrm{d}\sigma}{e^{2\sigma} - 1} \ .
\end{eqnarray}
We can further make use of the integral given by \cite{Grad}
\begin{eqnarray}
	\label{Int_0}
\int_{0}^{\infty}\frac{x^{\nu -1}\mathrm{d}x}{e^{\mu x} - 1}=\frac{1}{\mu ^{\nu }}\Gamma (\nu )\zeta (\nu ) \ ,
\end{eqnarray}
where $\Gamma (\nu )$ and $\zeta (\nu )$  correspond to the Gamma and Riemann zeta functions \cite{Abramo}, respectively. Thus, we find
\begin{eqnarray}
	\label{Ener_Cas}
\frac{\tilde{E_{C}}}{L^2} = \frac{1}{4\pi^{2} a^{3}}\left ( \frac{l}{a} \right )^{2(\epsilon-1)}\frac{(-1)^{\epsilon}}{(2\epsilon +1)}\frac{\Gamma (2\epsilon +2)\zeta (2\epsilon +2)}{2^{(2\epsilon +2)}} \ .
\end{eqnarray}
In this case, for instance, we have
\begin{itemize}
\item{For $\epsilon=2$}:
\end{itemize}
\begin{eqnarray}
\frac{\tilde{E_{C}}}{L^2} =\frac{\pi^{4}}{10080 a^{3}}\left ( \frac{l}{a} \right )^{2} \ .
\end{eqnarray}

\begin{itemize}
\item{For $\epsilon=3$}:
\end{itemize}
\begin{eqnarray}
\frac{\tilde{E_{C}}}{L^2} =-\frac{\pi^{6}}{13440 a^{3}}\left ( \frac{l}{a} \right )^{4} \ .
\end{eqnarray}
At this point we would like to emphasize that Eq. \eqref{Ener_Cas} has been derived considering $\epsilon\ge 2$. For this reason our results for LV Casimir energies cannot reduce to the corresponding ones obtained in \cite{Messias2019,Messias2020}  and \cite{Escobar1,Escobar2} by taking its limit $\epsilon=1$.

Unfortunately for the massive case, Eq. \eqref{energywithmassdpa} can only be given in terms of an infinite sum of modified Bessel functions as shown below; so let us evaluate its asymptotic limits for small and large values of the dimensionless parameter $ma$. In order to do this, we make the following change of variables $\xi^{2}=\sigma^{2}+(ma)^{2}$ and $\xi=mav$. Thus, it gives
\begin{eqnarray}
\label{exactdpa}
\frac{\tilde{E_{C}}}{L^2} = \frac{\left(am\right)^{2(\epsilon +1)}}{4\pi^{2} a^{3}}\left ( \frac{l}{a} \right )^{2(\epsilon-1)}\frac{(-1)^{\epsilon}}{(2\epsilon +1)}\int_{1}^{\infty}\frac{\left(v^{2}-1\right)^{\epsilon +\frac{1}{2}}\mathrm{d}v}{e^{2amv}-1} \ .
\end{eqnarray}
Knowing that the geometric series can be represented as
\begin{eqnarray}
	\label{geom_series}
\frac{1}{e^{2amv}-1}=\sum_{j=1}^{\infty}e^{-2amvj} \ ,
\end{eqnarray}
we obtain
\begin{eqnarray}
	\label{E_C_1}
\frac{\tilde{E_{C}}}{L^2} = \frac{\left(am\right)^{2(\epsilon +1)}}{4\pi^{2} a^{3}}\left ( \frac{l}{a} \right )^{2(\epsilon-1)}\frac{(-1)^{\epsilon}}{(2\epsilon +1)}\sum_{j=1}^{\infty}\int_{1}^{\infty}\left(v^{2}-1\right)^{\epsilon +\frac{1}{2}}e^{-2amvj}\mathrm{d}v \ .
\end{eqnarray}
The integral representation of the modified Bessel function, $K_\mu(z)$ \cite{Abramo}
\begin{eqnarray}
K_{\nu }(x) = \frac{\left ( \frac{x}{2} \right )^{\nu }\Gamma \left ( \frac{1}{2} \right )}{\Gamma \left ( \nu +\frac{1}{2} \right )}\int_{1}^{\infty}\mathrm{d}t\left ( t^{2}-1 \right )^{\nu -\frac{1}{2}}e^{-xt} \ ,
\end{eqnarray}
allows us to put the expression \eqref{E_C_1} in the form
\begin{eqnarray}
\frac{\tilde{E_{C}}}{L^2} = \frac{\left(am\right)^{\epsilon +1}\Gamma \left ( \epsilon +\frac{3}{2} \right )(-1)^{\epsilon }}{4{\left (\pi \right )^{\frac{5}{2}}} a^{3}(2\epsilon +1)}\left ( \frac{l}{a} \right )^{2(\epsilon-1)}\sum_{j=1}^{\infty}\frac{K_{\epsilon +1}(2amj)}{j^{\epsilon +1}} \ .
\end{eqnarray}

Let us now consider the expression above in two asymptotic regime cases:

$(i)$ The LV Casimir energy, $\tilde{E_{C}}$, for large values of $am>>1$, can be obtained by the using the asymptotic expression for the modified Bessel function for large arguments \cite{Abramo}:
\begin{eqnarray}
	\label{Asympt_BesselK}
K_{\nu}(z)\approx \sqrt{ \frac{\pi}{2z} }e^{-z} \ .
\end{eqnarray}
The dominant contribution is for $j=1$. So we get
\begin{eqnarray}
\frac{\tilde{E_{C}}}{L^2} \approx  \frac{\left(am\right)^{\epsilon+\frac{1}{2}}\Gamma \left ( \epsilon +\frac{3}{2} \right )(-1)^{\epsilon}}{8\pi^{2} a^{3}(2\epsilon +1)}\left ( \frac{l}{a} \right )^{2(\epsilon-1)}e^{-2am} \ .
\end{eqnarray}
We can observe that the Casimir energy decays exponentially. 

$(ii)$ For $am<<1$, it is better to consider Eq. \eqref{exactdpa}:
\begin{itemize}
\item{In the case $\epsilon=2$, the expression \eqref{exactdpa} becomes}
\end{itemize}
\begin{eqnarray}
\frac{\tilde{E_{C}}}{L^2} = \frac{\left(am\right)^{6}}{20\pi^{2} a^{3}}\left ( \frac{l}{a} \right )^{2}\int_{1}^{\infty}\frac{\left ( v^{2}-1\right )^{\frac{5}{2}}\mathrm{d}v}{e^{2amv}-1} \ .
\end{eqnarray}

We can approximate the integrand as shown below, and obtain a series expansion, i.e.,
\begin{eqnarray}
	\label{E_C_1}
\frac{\tilde{E_{C}}}{L^2} &\approx& \frac{\left(am\right)^{6}}{20\pi^{2} a^{3}}\left ( \frac{l}{a} \right )^{2}\int_{1}^{\infty}\frac{\left ( v^{5}-\frac{5}{2}v^{3}+\frac{15}{8}v\right )\mathrm{d}v}{e^{2amv}-1}\nonumber\\&\approx&\frac{1}{10080\pi^{2} a^{3}}\left ( \frac{l}{a} \right )^{2}\left [\pi^{6}-\frac{21}{4}\pi^{4}(am)^{2}+\frac{315}{8}\pi^{2}(am)^{4}  \right ] \ .
\end{eqnarray}

The LV Casimir pressure can be obtained by the standard procedure:
\begin{eqnarray}
		\label{Pre}
	{\tilde P}_C(a)=-\frac1{L^2}\frac{\partial{\tilde{E_{C}}}}	{\partial a} \  .
\end{eqnarray}

Taking the approximated expression \eqref{E_C_1}, we get:
\begin{eqnarray}
	{\tilde P}_C(a)=\frac 1{80640a^4}\left ( \frac{l}{a} \right )^{2}\left[40\pi^4-126\pi^2(am)^2+315(am)^4\right].
\end{eqnarray}

\begin{itemize}
\item{In the case $\epsilon=3$, the expression \eqref{exactdpa} becomes}:
\end{itemize}
\begin{eqnarray}
\frac{\tilde{E_{C}}}{L^2} = -\frac{\left(am\right)^{8}}{28\pi^{2} a^{3}}\left ( \frac{l}{a} \right )^{4}\int_{1}^{\infty}\frac{\left ( v^{2}-1\right )^{\frac{7}{2}}\mathrm{d}v}{e^{2amv}-1} \ .
\end{eqnarray}

Adopting the same procedure to approximate the integrand as above, the series expansion is
\begin{eqnarray}
\frac{\tilde{E_{C}}}{L^2} &\approx& -\frac{\left(am\right)^{8}}{28\pi^{2} a^{3}}\left ( \frac{l}{a} \right )^{4}\int_{1}^{\infty}\frac{\left ( v^{7}-\frac{7}{2}v^{5}+\frac{35}{8}v^{3}-\frac{35}{16}v\right )\mathrm{d}v}{e^{2amv}-1}\nonumber\\&\approx&-\frac{1}{13440\pi^{2} a^{3}}\left ( \frac{l}{a} \right )^{4}\left [\pi^{8}-\frac{10}{3}\pi^{6}(am)^{2}+\frac{35}{4}\pi^{4}(am)^{4}  \right ] \ .
\end{eqnarray}
Taking \eqref{Pre} the corresponding LV Casimir pressure is
\begin{eqnarray}
	{\tilde P}_C(a)=-\frac 1{161280a^4}\left ( \frac{l}{a} \right )^{4}\left[84\pi^6-200\pi^4(am)^2+315\pi^2(am)^4\right].
\end{eqnarray}

\begin{figure}[h]
\subfigure[For $\epsilon=2$\label{Dpa2}]{\includegraphics[scale=0.35]{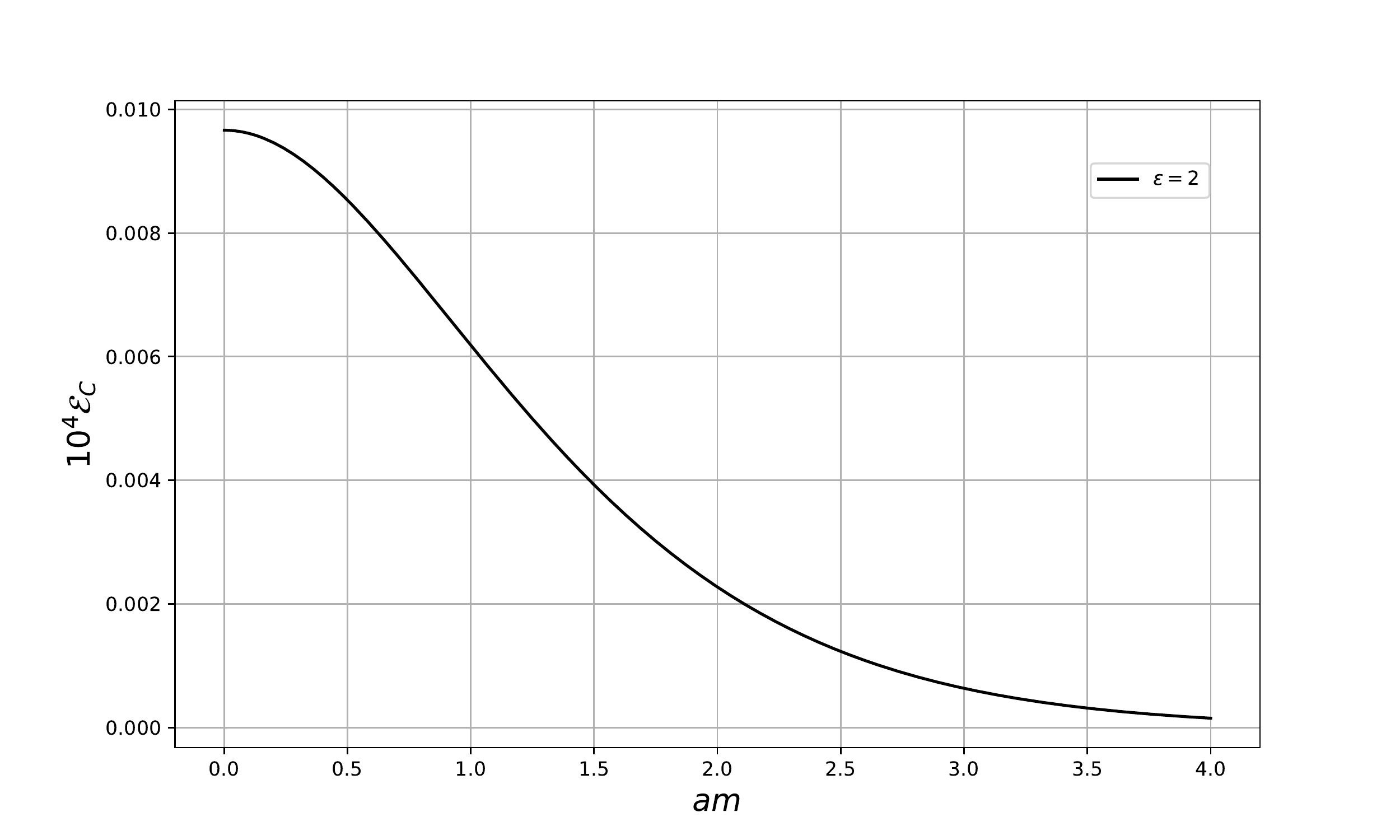}}
\subfigure[For $\epsilon=3$\label{Dpa3}]{\includegraphics[scale=0.35]{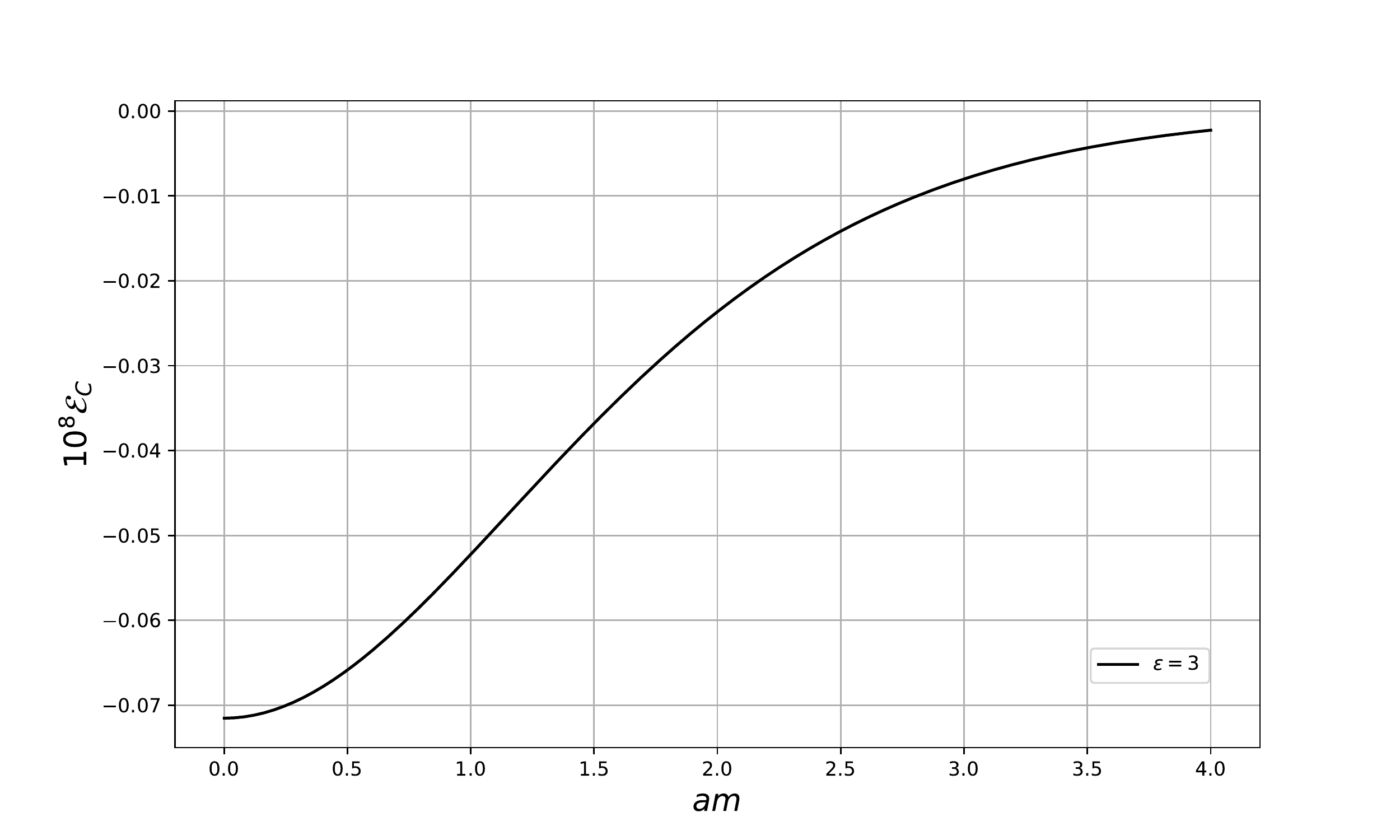}}
\caption{The Casimir energy per unit area multiplied by $a^3$ as function of $ma$ in case $u^{\mu}=(0,1,0,0)$, and the field obeying Dirichlet boundary condition, for $\epsilon=2$ in the left panel, and $\epsilon=3$ in the right panel. In both plots we have considered $\frac{l}{a}=0.01$.}
\label{fig2}
\end{figure}

In Fig \ref{fig2} we present the behavior of the Casimir energy per unit area multiplied by $a^3$, $\varepsilon _{c}=\frac{\tilde{E_{C}}}{L^2}a^{3}$, as a function of $ma$, considering as only an illustrative example $\frac l a=0.01$, for two distinct values of $\epsilon$. In the plot on the left we consider $\epsilon=2$, while on the right we consider $\epsilon=3$. 

\subsubsection{ Vector perpendicular to the plates}%%%%%%%%%%%%%%%%%%%%%%%%%%%%%%%%%%%%%%%%%%%%%%%%%%%%%%%%%%%%%%%%%

Let us now consider that the four-vector $u^{\mu}$ is perpendicular to the plates, i.e.,
\begin{eqnarray}
u^{\mu}=(0,0,0,1) \ .
\end{eqnarray}

The corresponding dispersion relation is
\begin{eqnarray}
\omega_{\textbf{k},n}^2 = k_{x}^2 + k_{y}^2 + \left(\frac{n\pi}{a}\right)^2 + l^{2(\epsilon-1)}(-1)^{\epsilon}\left(\frac{n\pi}{a}\right)^{2\epsilon} + m^2 \ .
\end{eqnarray}

For this case, the Hamiltonian operator, $\hat{H}$, has the same structure as \eqref{Hamilt}, consequently the vacuum energy is given by 
\begin{eqnarray}
E_{0} = \bra{0}\hat{H}\ket{0} = \frac{L^2}{8 \pi^2}\int \mathrm{d}^2\textbf{k} 
\sum_{n=1}^{\infty}\omega_{\textbf{k},n} \  .
\end{eqnarray}

Performing a change of coordinates $(k_{x},k_{y})$ to polar ones, $(k,\theta)$, and also a change of variable $u=ak$, we obtain
\begin{eqnarray}
	\label{E_C_2}
E_{0}=\frac{L^2}{8 \pi^2a^{3}}\int_{0}^{2\pi}d\theta\int_{0}^{\infty}udu
\sum_{n=1}^{\infty}\left [ u^{2}+(n\pi)^{2}+(ma)^{2}+\left ( \frac{l}{a} \right )^{2(\epsilon-1)}(-1)^{\epsilon}(n\pi)^{2\epsilon }\right ]^{\frac{1}{2}} \  .
\end{eqnarray}
For this case the integral over the angular variable is trivial. However, to obtain the Casimir energy we have to develop the summation over $n$. In this sense, by using the Abel-Plana formula, Eq. \eqref{Abel}, we have not found in the literature a very enlightening result 
for the integral over the variable $t$, for any value of $\epsilon$. So, by adopting an analogous procedure as in the last subsection, we perform an expansion in the parameter $\frac{l}{a}\ll 1$ in the integrand of \eqref{E_C_2}. Doing this, the leading term in the approximated expression for $E_0$ is given by
\begin{eqnarray}
E_{0} &\approx& \frac{L^2}{8\pi^2a^{3}}\int_{0}^{2\pi}d\theta\int_{0}^{\infty}udu\sum_{n=1}^{\infty}\Bigg\{\left[u^{2}+(n\pi)^{2}+(ma)^{2}\right ]^{\frac{1}{2}}\nonumber\\ 
&+&\frac{1}{2}\left(\frac{l}{a}\right)^{2(\epsilon-1)}(-1)^{\epsilon}(n\pi)^{2\epsilon }\left [ u^{2}+(n\pi)^{2}+(ma)^{2}\right ]^{-\frac{1}{2}}\Bigg\}\ .
\end{eqnarray}
Again, the first term is associated with the Casimir energy without Lorentz violation. Thus, after integration over the angular coordinate, the second term becomes
\begin{eqnarray}
\label{energywhioutviolationdpe}
\tilde{E_{0}}=\frac{L^2(-1)^{\epsilon }}{8 \pi a^{3}}\left ( \frac{l}{a} \right )^{2(\epsilon-1)}\int_{0}^{\infty}u\mathrm{d}u
\sum_{n=1}^{\infty}(n\pi)^{2\epsilon }\left[u^{2}+(n\pi)^{2}+(ma)^{2}\right ]^{-\frac{1}{2}} \  .
\end{eqnarray}
Using the summation formula \eqref{Abel}, we find
\begin{eqnarray}
\tilde{E_{0}}=\frac{L^2(-1)^{\epsilon }}{8 \pi a^{3}}\left ( \frac{l}{a} \right )^{2(\epsilon-1)}\int_{0}^{\infty}u\mathrm{d}u\Bigg\{-\frac{1}{2}F(0)+\int_{0}^{\infty} 
F(t)\mathrm{d}t+i\int_{0}^{\infty}\frac{F(it)-F(-it)}{e^{2\pi t}-1}\mathrm{d}t\Bigg\}\   ,
\end{eqnarray}
where 
\begin{eqnarray}
F(n) = (n\pi)^{2\epsilon }\left[u^{2}+(n\pi)^{2}+(ma)^{2}\right ]^{-\frac{1}{2}} \  .
\end{eqnarray}
Discarding the divergent contributions coming from the first two terms, the LV Casimir energy per unit area of the plates is given by
\begin{eqnarray}
	\label{E_C_4}
\frac{\tilde{E_{C}}}{L^2}&=&\frac{i}{8 \pi a^{3}}\left ( \frac{l}{a} \right )^{2(\epsilon-1)}\int_{0}^{\infty}u\mathrm{d}u\nonumber\\&\times&\int_{0}^{\infty}\mathrm{d}v \frac{v^{2\epsilon }\Bigg\{[u^2 + 
(iv)^{2}+(ma)^{2}]^{-1/2} - [u^2 + 
(-iv)^{2}+(ma)^{2}]^{-1/2}\Bigg\}}{e^{2v} - 1} \  ,
\end{eqnarray}
where we have performed a change of variable $t\pi = v$.

Again, analyzing the integral in the variable $v$ over the two intervals $v<\left [ u^2 +(ma)^{2} \right ]^{1/2}$ and $v>\left [ u^2 +(ma)^{2} \right ]^{1/2}$, we obtain
\begin{eqnarray}
\frac{\tilde{E_{C}}}{L^2} = \frac{1}{4\pi^{2} a^{3}}\left ( \frac{l}{a} \right )^{2(\epsilon-1)}\int_{0}^{\infty}u\mathrm{d}u\int_{\left [ u^{2}+(ma)^{2} \right ]^{\frac{1}{2}}}^{\infty}\mathrm{d}v\frac{v^{2\epsilon }[v^2-(u^{2}+(ma)^{2})]^{-1/2}}{e^{2v} - 1} \ .
\label{CEPer}
\end{eqnarray}
Next, performing a changing of variable $\rho^{2} =v^{2}-(u^{2}+(ma)^{2})$ and also a change of coordinates in the plane  $(u,\rho)$ to polar ones, we are able to re-write Eq. \eqref{CEPer} in the form
\begin{eqnarray}
\label{energywithmassdpe}
\frac{\tilde{E_{C}}}{L^2} = \frac{1}{4\pi^{2} a^{3}}\left ( \frac{l}{a} \right )^{2(\epsilon-1)}\int_{0}^{\infty}\frac{\left [ \sigma ^{2}+(ma)^{2} \right ]^{\epsilon -\frac{1}{2}}\sigma^{2}\mathrm{d}\sigma}{e^{2(\sigma^{2}+(ma)^{2})^{\frac{1}{2}}} - 1} \ .
\end{eqnarray}

For the massless scalar field case we have
\begin{eqnarray}
\frac{\tilde{E_{C}}}{L^2} = \frac{1}{4\pi^{2} a^{3}}\left ( \frac{l}{a} \right )^{2(\epsilon-1)}\int_{0}^{\infty}\frac{\sigma^{2\epsilon+1}\mathrm{d}\sigma}{e^{2\sigma} - 1} \ .
\end{eqnarray}
Consequently, by using \eqref{Int_0}, we obtain
\begin{eqnarray}
\frac{\tilde{E_{C}}}{L^2} = \frac{1}{4\pi^{2} a^{3}}\left ( \frac{l}{a} \right )^{2(\epsilon-1)}\frac{\Gamma (2\epsilon +2)\zeta (2\epsilon +2)}{2^{(2\epsilon +2)}} \ .
\end{eqnarray}
Thus, we can analyze two cases:
\begin{itemize}
\item{For $\epsilon=2$}:
\end{itemize}
\begin{eqnarray}
\frac{\tilde{E_{C}}}{L^2} =\frac{\pi^{4}}{2016 a^{3}}\left ( \frac{l}{a} \right )^{2} \ .
\end{eqnarray}

\begin{itemize}
\item{For $\epsilon=3$}:
\end{itemize}
\begin{eqnarray}
\frac{\tilde{E_{C}}}{L^2} =\frac{\pi^{6}}{1920 a^{3}}\left ( \frac{l}{a} \right )^{4} \ .
\end{eqnarray}

The integral in \eqref{energywithmassdpe} can only be expressed in terms of an infinite series in  modified Bessel functions for $m\neq0$; so let us evaluate its asymptotic limits. To do this, we make the following change of variables $\xi^{2}=\sigma^{2}+(ma)^{2}$ and $\xi=mav$. So, we have
\begin{eqnarray}
\label{exactdper}
\frac{\tilde{E_{C}}}{L^2} = \frac{\left(am\right)^{2(\epsilon +1)}}{4\pi^{2} a^{3}}\left ( \frac{l}{a} \right )^{2(\epsilon-1)}\int_{1}^{\infty}\frac{v^{2\epsilon }\left(v^{2}-1\right)^{\frac{1}{2}}\mathrm{d}v}{e^{2amv}-1} \ .
\end{eqnarray}

Expressing the denominator in terms of a geometric series as shown in  \eqref{geom_series}, we can re-write \eqref{exactdper} as
\begin{eqnarray}
\frac{\tilde{E_{C}}}{L^2} = \frac{\left(am\right)^{2(\epsilon +1)}}{4\pi^{2} a^{3}}\left ( \frac{l}{a} \right )^{2(\epsilon-1)}\sum_{j=1}^{\infty}\int_{1}^{\infty}v^{2\epsilon }\left(v^{2}-1\right)^{\frac{1}{2}}e^{-2amvj}\mathrm{d}v \ .
\label{Energy}
\end{eqnarray}

The use of the identity,
\begin{eqnarray}
\frac{1}{\left ( 2mj \right )^{2\epsilon }}\frac{\mathrm{d^{2\epsilon }} \left ( e^{-2amvj} \right )}{\mathrm{d} a^{2\epsilon }}=v^{2\epsilon }e^{-2amvj} \ ,
\end{eqnarray}
allows us to put Eq. \eqref{Energy} in the form
\begin{eqnarray}
\frac{\tilde{E_{C}}}{L^2} = \frac{\left(am\right)^{2(\epsilon +1)}}{4\pi^{2} a^{3}}\left ( \frac{l}{a} \right )^{2(\epsilon-1)}\frac{1}{\left ( 2m \right )^{2\epsilon }}\sum_{j=1}^{\infty}\frac{1}{j^{2\epsilon }}\frac{\mathrm{d^{2\epsilon }} }{\mathrm{d} a^{2\epsilon }}\int_{1}^{\infty}\left(v^{2}-1\right)^{\frac{1}{2}}e^{-2amvj}\mathrm{d}v \ .
\end{eqnarray}
Again, by making use of the integral representation for the modified Bessel function, the above expression reads
\begin{eqnarray}
\frac{\tilde{E_{C}}}{L^2} = \frac{\left(am\right)^{2(\epsilon +1)}}{4\pi^{2} a^{3}}\left ( \frac{l}{a} \right )^{2(\epsilon-1)}\frac{1}{(2m)^{2\epsilon +1}}\sum_{j=1}^{\infty}\frac{1}{j^{2\epsilon +1}}\frac{\mathrm{d^{2\epsilon }} }{\mathrm{d} a^{2\epsilon }}\left ( \frac{K_{1}(2amj)}{a} \right ) \ .
\end{eqnarray}

Let us now consider the asymptotic limits of the above result for $am>>1$ and for $am<<1$:

(i) For large values of $am>>1$, and using the asymptotic expression \eqref{Asympt_BesselK} for the modified Bessel function, the dominant term provides, 
\begin{eqnarray}
\frac{\tilde{E_{C}}}{L^2} \approx  \frac{\left(am\right)^{2(\epsilon+1)}}{8\left ( \pi \right )^{\frac{3}{2}} a^{3}m^{\frac{1}{2}}(2m)^{2\epsilon +1}}\left ( \frac{l}{a} \right )^{2(\epsilon-1)}\frac{\mathrm{d^{2\epsilon }} }{\mathrm{d} a^{2\epsilon }}\left ( \frac{e^{-2am}}{a^{\frac{3}{2}}} \right ) \ .
\end{eqnarray}
In this sense, we have
\begin{itemize}
\item{For case $\epsilon=2$}:
\end{itemize}
\begin{eqnarray}
\frac{\tilde{E_{C}}}{L^2} \approx \frac{\left(am\right)^{\frac{9}{2}}}{16\left ( \pi \right )^{\frac{3}{2}}a^{3}}\left ( \frac{l}{a} \right )^{2}e^{-2am} \ .
\end{eqnarray}

\begin{itemize}
\item{For case $\epsilon=3$}:
\end{itemize}
\begin{eqnarray}
\frac{\tilde{E_{C}}}{L^2} \approx \frac{\left(am\right)^{\frac{13}{2}}}{16\left ( \pi \right )^{\frac{3}{2}}a^{3}}\left ( \frac{l}{a} \right )^{4}e^{-2am} \ .
\end{eqnarray}

(ii) For $am<<1$, we have to take the integral representation \eqref{exactdper}
\begin{itemize}
\item{In the case $\epsilon=2$, Eq. \eqref{exactdper} becomes}:
\end{itemize}
\begin{eqnarray}
\frac{\tilde{E_{C}}}{L^2} = \frac{\left(am\right)^{6}}{4\pi^{2} a^{3}}\left ( \frac{l}{a} \right )^{2}\int_{1}^{\infty}\frac{v^{4}\left ( v^{2}-1\right )^{\frac{1}{2}}\mathrm{d}v}{e^{2amv}-1} \ .
\end{eqnarray}
By approximating the integrand as shown below, we obtain 
\begin{eqnarray}
\frac{\tilde{E_{C}}}{L^2} &\approx& \frac{\left(am\right)^{6}}{4\pi^{2} a^{3}}\left ( \frac{l}{a} \right )^{2}\int_{1}^{\infty}\frac{\left ( v^{5}-\frac{1}{2}v^{3}-\frac{1}{8}v\right )\mathrm{d}v}{e^{2amv}-1}\nonumber\\&\approx&\frac{1}{2016\pi^{2} a^{3}}\left ( \frac{l}{a} \right )^{2}\left [\pi^{6}-\frac{21}{20}\pi^{4}(am)^{2}-\frac{21}{8}\pi^{2}(am)^{4}  \right ] \ .
\end{eqnarray}
For this case the pressure is given by
\begin{eqnarray}
	{\tilde P}_C(a)=\frac 1{80640a^4}\left ( \frac{l}{a} \right )^{2}\left[200\pi^4-126\pi^2(am)^2+105(am)^4\right].
\end{eqnarray}

\begin{itemize}
\item{In the case $\epsilon=3$, the expression \eqref{exactdper} becomes}:
\end{itemize}
\begin{eqnarray}
\frac{\tilde{E_{C}}}{L^2} = \frac{\left(am\right)^{8}}{4\pi^{2} a^{3}}\left ( \frac{l}{a} \right )^{4}\int_{1}^{\infty}\frac{v^{6}\left ( v^{2}-1\right )^{\frac{1}{2}}\mathrm{d}v}{e^{2amv}-1} \ .
\end{eqnarray}

So, the series expansion is now given by
\begin{eqnarray}
\frac{\tilde{E_{C}}}{L^2} &\approx& \frac{\left(am\right)^{8}}{4\pi^{2} a^{3}}\left ( \frac{l}{a} \right )^{4}\int_{1}^{\infty}\frac{\left ( v^{7}-\frac{1}{2}v^{5}-\frac{1}{8}v^{3}-\frac{1}{16}v\right )\mathrm{d}v}{e^{2amv}-1}\nonumber\\&\approx&\frac{1}{1920\pi^{2} a^{3}}\left ( \frac{l}{a} \right )^{4}\left [\pi^{8}-\frac{10}{21}\pi^{6}(am)^{2}-\frac{1}{4}\pi^{4}(am)^{4}  \right ] \ .
\end{eqnarray}
Consequently, the pressure is found to be
\begin{eqnarray}
	{\tilde P}_C(a)=\frac 1{161280a^4}\left ( \frac{l}{a} \right )^{4}\left[588\pi^6-200\pi^4(am)^2+63\pi^2(am)^4\right].
\end{eqnarray}

\begin{figure}[h]
\subfigure[For $\epsilon=2$\label{Dper2}]{\includegraphics[scale=0.35]{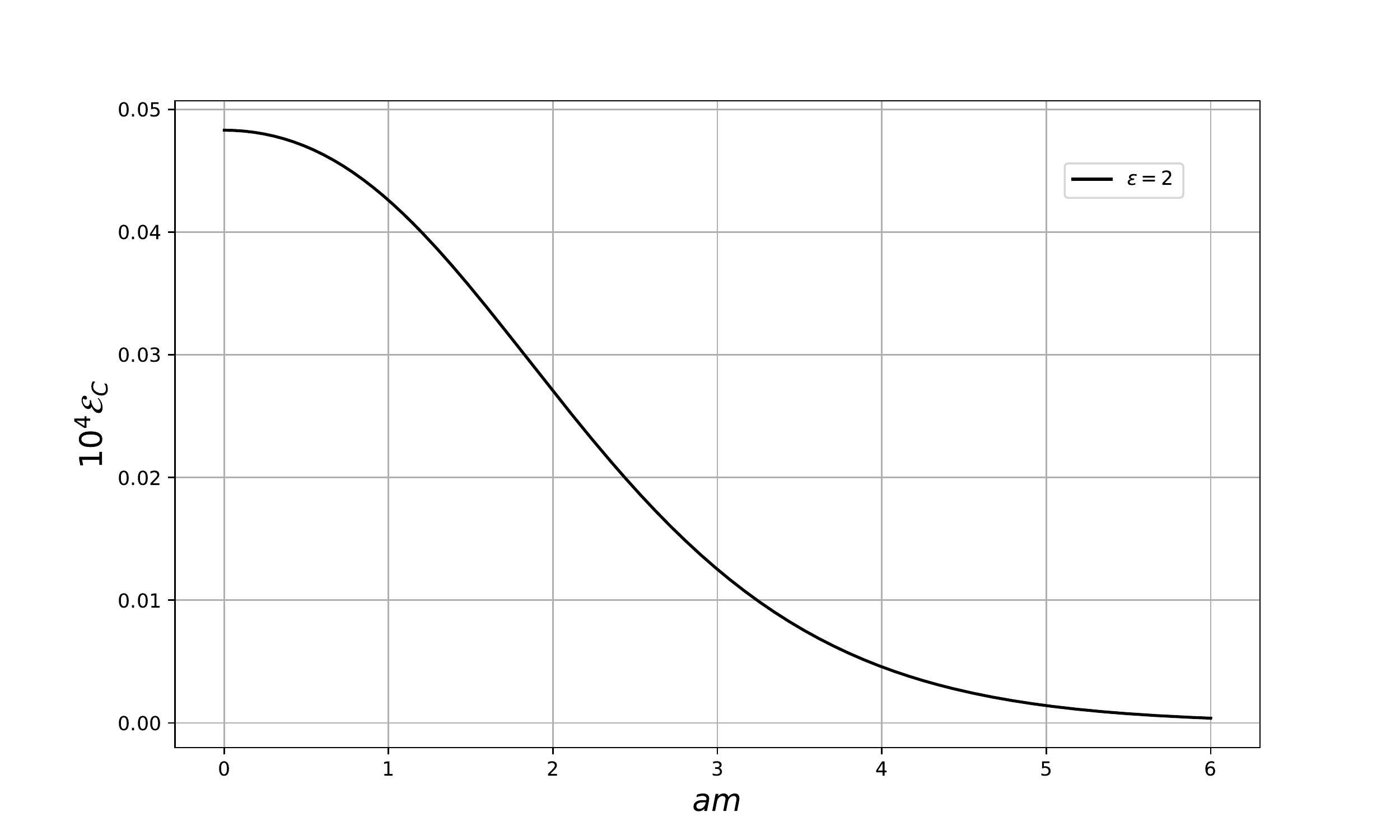}}
\subfigure[For $\epsilon=3$\label{Dper3}]{\includegraphics[scale=0.35]{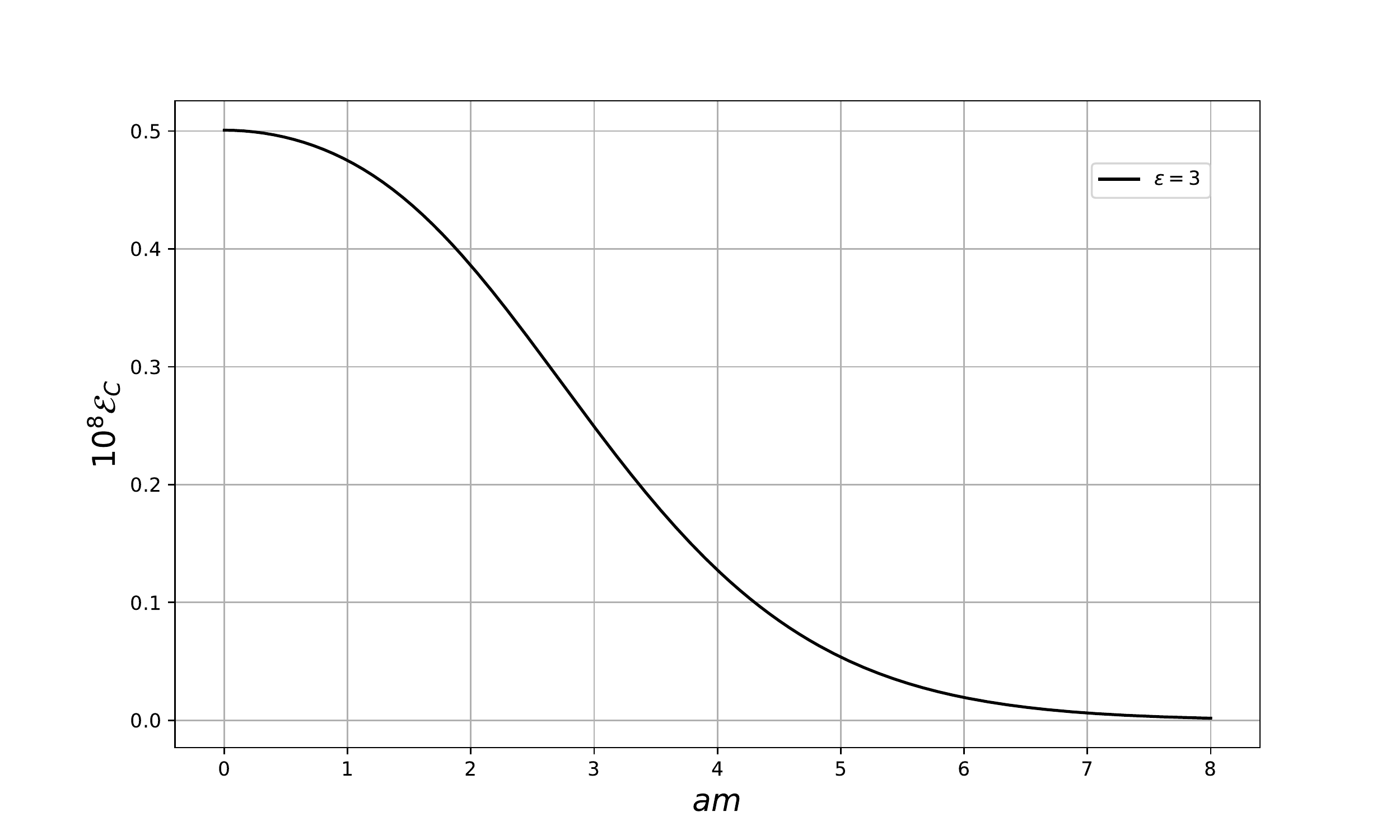}}
\caption{The Casimir energy per unit area multiplied by $a^3$, as function of $am$ in the case $u^{\mu}=(0,0,0,1)$ for Dirichlet condition applied on the plates. In the left panel we have adopted $\epsilon=2$, while in the right panel, $\epsilon=3$. For both cases we assume $\frac{l}{a}=0.01$.}
\label{fig3}
\end{figure}

In Fig. \ref{fig3} we present the behavior of the Casimir energy per unit area multiplied by $a^3$, $\varepsilon _{c}=\frac{\tilde{E_{C}}}{L^2}a^{3}$, as function of $ma$, considering as an illustrative example $\frac l a=0.01$, for two distinct values of $\epsilon$. In the left panel we adopted $\epsilon=2$, and in the right panel $\epsilon=3$.

\subsection{Neumann condition}%%%%%%%%%%%%%%%%%%%%%%%%%%%%%%%%%%%%%%%%%%%%%%%%%%%%%%%%%%%%%%%%%%%%%%%

We turn now to Neumann boundary condition. Thus, the field operator solution of Eq. \eqref{kgmodific} which obey the boundary condition below, 
\begin{eqnarray}
\frac{\partial \phi(\textbf{x})}{\partial z}\Big |_{z=0} = 
\frac{\partial \phi(\textbf{x})}{\partial z}\Big |_{z=a} = 0 \ ,
\end{eqnarray}
has been given in \cite{Messias2017}. It reads, 
\begin{eqnarray}
	\label{Field_N}
\hat{\phi}(x) = \int \mathrm{d}^2\textbf{k} \sum_{n=0}^{\infty} c_{n} 
\cos\left(\dfrac{n\pi}{a}z\right)[\hat{a}_{\textbf{k},n}e^{-i k x} + \hat{a}^{\dagger}_{\textbf{k},n}e^{i k x}] \  ,
\end{eqnarray}
where
\begin{equation}
c_{n} = \begin{cases}
\dfrac{1}{\sqrt{2(2\pi)^2 a \  \omega_{\textbf{k},n}}} & \text{for $n = 0$}, \\
\dfrac{1}{\sqrt{(2\pi)^2 a \ \omega_{\textbf{k},n}}} & \text{for $n \ge 0$}.
\end{cases}
\end{equation} 
Note that, although the field operator is different from the one presented in the previous subsection, \eqref{Field_Diri}, the Hamiltonian operator and the dispersion relations remain the same as to the Dirichlet boundary condition for each choice of the space-like four-vector. So, for this reason we decided do not present all the calculations for this case.

\subsection{Mixed boundary condition}%%%%%%%%%%%%%%%%%%%%%%%%%%%%%%%%%%%%%%%%%%%%%%%%%%%%%%%%%%%%%%%%%%%%
\label{Mixed}
Now, let us consider that the scalar quantum field solution of Eq. \eqref{kgmodific} obeys Dirichlet boundary condition on one plate and Neumann on the other. In this case, we have two different configurations for the scalar quantum field:
\begin{itemize}
\item First configuration,
\begin{eqnarray}
\phi(z=0) = \frac{\partial\phi(\textbf{x})}{\partial z} |_{z=a}=0 \ .
\end{eqnarray}
\item Second configuration,
\begin{eqnarray}
\frac{\partial\phi(\textbf{x})}{\partial z} |_{z=0} = \phi(z=a)  =0 \ .
\end{eqnarray}
\end{itemize}

For this case, the field operators are given by
\begin{eqnarray}
	\label{Mixed_1}
\hat{\phi}_{(i)}(x) = \int \mathrm{d}^2 \textbf{k} \sum_{n = 0}^{\infty} \frac{1}
{\sqrt{(2 \pi)^2 a  \ \omega_{\textbf{k},n}}} \sin\left((n+1/2)\frac{ \pi}{a}z\right)
[ \hat{a}_{\textbf{k},n}e^{-ikx} + \hat{a}_{\textbf{k},n}^{\dagger} e^{ikx}]
\end{eqnarray}
for the first configuration and
\begin{eqnarray}
	\label{Mixed_2}
\hat{\phi}_{(ii)}(x) = \int \mathrm{d}^2 \textbf{k} \sum_{n = 0}^{\infty} 
\frac{1}{\sqrt{(2 \pi)^2 a \ \omega_{\textbf{k},n}}} \cos\left((n+1/2)\frac{ \pi}
{a}z\right)[ \hat{a}_{\textbf{k},n}e^{-ikx} + \hat{a}_{\textbf{k},n}^{\dagger} e^{ikx}] \  , 
\end{eqnarray}
for the second configuration. However, the field operators $\hat{\phi}_{1}$ and $\hat{\phi}_{2}$ provide the same hamiltonian operator and dispersion relations.

\subsubsection{Vector parallel to the plates}%%%%%%%%%%%%%%%%%%%%%%%%%%%%%%%%%%%%%%%%%%%%%%%%%%%%%%%%%%%%%%%%%%%%

Considering $u^{\mu}=(0,1,0,0)$, the field presents the following dispersion relation,
\begin{eqnarray}
\omega_{\textbf{k},n}^2 = k_{x}^2 + k_{y}^2 + l^{2(\epsilon-1)}(-1)^{\epsilon}k_{x}^{2\epsilon} + \left [ (n+1/2)\frac{\pi}{a} \right ]^{2} + m^2 \ .
\end{eqnarray}
Note that both field operators, $\hat{\phi}_{(i)}(x)$ and $\hat{\phi}_{(ii)}(x)$, provide the same 
Hamiltonian operator, i.e.,
\begin{eqnarray}
\hat{H} = \frac{1}{2}\int\mathrm{d}^2\textbf{k} \sum_{n=0}^{\infty}\omega_{\textbf{k},n}\left[2 \hat{a}^{\dagger}_{\textbf{k,n}} \hat{a}_{\textbf{k,n}} + 
\frac{L^2}{(2\pi)^2}\right] \   .
\end{eqnarray}
Consequently, the corresponding vacuum energy is expressed as
\begin{eqnarray}
E_{0} = \bra{0}\hat{H}\ket{0} = \frac{L^2}{8 \pi^2}\int \mathrm{d}^2\textbf{k} 
\sum_{n=0}^{\infty}\omega_{\textbf{k},n} \  .
\end{eqnarray}

By performing a change of coordinates in the plane $(k_{x},k_{y})$ to the polar ones, making the change of variable $u=ak$ and doing an expansion in the dimensionless parameter $\frac{l}{a}<<1$, up to the first order in this parameter, we find
\begin{eqnarray}
E_{0}&\approx&\frac{L^2}{8\pi^2a^{3}}\int_{0}^{2\pi}\mathrm{d}\theta\int_{0}^{\infty}u\mathrm{d}u\sum_{n=0}^{\infty}\Bigg\{\left[u^{2}+\left [ (n+1/2)\pi \right ]^{2}+(ma)^{2}\right ]^{\frac{1}{2}}\nonumber\\ &+&\frac{1}{2}\left(\frac{l}{a}\right)^{2(\epsilon-1)}(-1)^{\epsilon}u^{2\epsilon}\cos^{2\epsilon}{\theta}\left [ u^{2}+\left [ (n+1/2)\pi \right ]^{2}+(ma)^{2}\right ]^{-\frac{1}{2}}\Bigg\}\ .
\end{eqnarray}

Note that the first term on the right hand side of the above expression is associated with the vacuum energy without Lorentz violation. An integral representation for the  corresponding Casmir energy by unity area for this case \cite{Messias2017}, reads
\begin{eqnarray}
	\label{int4}
	\frac{E_C}{L^2}=\frac{am^4}{6\pi^{2}}\int_{1}^{\infty}\frac{(v^2-1)^{\frac{3}{2}}}{e^{2amv}+1}{dv}.
\end{eqnarray}
 As to the second term, after integration over the angular coordinate, we have
\begin{eqnarray}
\label{energywhioutviolationmpa}
\tilde{E_{0}}=\frac{L^2}{8 \pi a^{3}}\left ( \frac{l}{a} \right )^{2(\epsilon-1)}\frac{(-1)^{\epsilon}(2\epsilon-1)!!}{(2\epsilon)!!}\int_{0}^{\infty}u^{^{(2\epsilon+1)}}\mathrm{d}u
\sum_{n=0}^{\infty}\left[u^{2}+\left [ (n+1/2)\pi \right ]^{2}+(ma)^{2}\right ]^{-\frac{1}{2}} \  .
\end{eqnarray}

In order to develop the summation over half-integer number, we will use the Abel-Plana formula below \cite{Bordag:2009zzd}:
\begin{eqnarray}
\label{Abel-half-integer}
\sum_{n=0}^{\infty}F\left ( n+\frac{1}{2} \right ) = \int_{0}^{\infty} F(t)\mathrm{d}t - 
i \int_{0}^{\infty}\frac{\mathrm{d}t}{e^{2\pi t}+1}[ F(it) - F(-i t)] \  .
\end{eqnarray}
Then the expression \eqref{energywhioutviolationmpa} becomes
\begin{eqnarray}
\label{energywhithabelmpa}
\tilde{E_{0}} = \frac{L^2}{8 \pi a^{3}}\left ( \frac{l}{a} \right )^{2(\epsilon-1)}\frac{(-1)^{\epsilon}(2\epsilon-1)!!}{(2\epsilon)!!}\int_{0}^{\infty}u^{^{(2\epsilon+1)}}\mathrm{d}u\Bigg\{\int_{0}^{\infty} 
F(t)\mathrm{d}t - i\int_{0}^{\infty}\frac{F(it)-F(-it)}{e^{2\pi t}+1}\mathrm{d}t\Bigg\}\   ,
\end{eqnarray}
where 
\begin{eqnarray}
F\left ( n+\frac{1}{2} \right ) = \left[u^{2}+\left [ (n+1/2)\pi \right ]^{2}+(ma)^{2}\right ]^{-\frac{1}{2}} \  . 
\end{eqnarray}
The first term in the right hand side of \eqref{energywhithabelmpa} refers to the free vacuum energy, so it is discarded in the renormalization process. Then, the LV Casimir energy will be given by
\begin{eqnarray}
\frac{\tilde{E_{C}}}{L^2} &=& -\frac{i}{8 \pi a^{3}}\left ( \frac{l}{a} \right )^{2(\epsilon-1)}\frac{(-1)^{\epsilon}(2\epsilon-1)!!}{(2\epsilon)!!}\int_{0}^{\infty}u^{^{(2\epsilon+1)}}\mathrm{d}u\nonumber\\&\times&\int_{0}^{\infty}\mathrm{d}t \frac{[u^2 + 
(it\pi)^{2}+(ma)^{2}]^{-1/2} -[u^2 + 
(-it\pi)^{2}+(ma)^{2}]^{-1/2}}{e^{2 \pi t} + 1} \  .
\label{VEMix}
\end{eqnarray}

Performing a change of variable, where $t\pi = v$, we can re-write Eq. \eqref{VEMix} in the form
\begin{eqnarray}
\frac{\tilde{E_{C}}}{L^2}&=&-\frac{i}{8 \pi^{2} a^{3}}\left ( \frac{l}{a} \right )^{2(\epsilon-1)}\frac{(-1)^{\epsilon}(2\epsilon-1)!!}{(2\epsilon)!!}\int_{0}^{\infty}u^{^{(2\epsilon+1)}}\mathrm{d}u\nonumber\\&\times&\int_{0}^{\infty}\mathrm{d}v\frac{[u^2 +(ma)^{2}+(iv)^2]^{-1/2} - 
[u^2 +(ma)^{2}+(-iv)^2]^{-1/2}}{e^{2v} + 1} \ . 
\end{eqnarray}
Now, carrying out the same analysis as in the previous cases, for the interval of the integral in $v$, we obtain 
\begin{eqnarray}
\frac{\tilde{E_{C}}}{L^2} &=& -\frac{1}{4\pi^{2} a^{3}}\left ( \frac{l}{a} \right )^{2(\epsilon-1)}\frac{(-1)^{\epsilon}(2\epsilon-1)!!}{(2\epsilon)!!}\int_{0}^{\infty}u^{^{(2\epsilon+1)}}\mathrm{d}u\nonumber\\&\times&\int_{\left [ u^{2}+(ma)^{2} \right ]^{\frac{1}{2}}}^{\infty}\mathrm{d}v\frac{[v^2-(u^{2}+(ma)^{2})]^{-1/2}}{e^{2v} + 1} \ .
\end{eqnarray}
Additionally, the new change of variable $\rho^{2} =v^{2}-(u^{2}+(ma)^{2})$ provides
\begin{eqnarray}
\frac{\tilde{E_{C}}}{L^2} &=& -\frac{1}{4\pi^{2} a^{3}}\left ( \frac{l}{a} \right )^{2(\epsilon-1)}\frac{(-1)^{\epsilon}(2\epsilon-1)!!}{(2\epsilon)!!}\int_{0}^{\infty}u^{^{(2\epsilon+1)}}\mathrm{d}u\nonumber\\&\times&\int_{0}^{\infty}\frac{\mathrm{d}\rho }{\left [ \rho^{2}+u^{2}+(ma)^{2} \right ]^{\frac{1}{2}}\left [e^{2( \rho^{2}+u^{2}+(ma)^{2})^{\frac{1}{2}}} + 1\right ]} \ .
\end{eqnarray}
Finally, by changing the coordinates in the plane  $(u,\rho)$ to polar ones, we find
\begin{eqnarray}
\label{energywithmassmpa}
\frac{\tilde{E_{C}}}{L^2} = -\frac{1}{4\pi^{2} a^{3}}\left ( \frac{l}{a} \right )^{2(\epsilon-1)}\frac{(-1)^{\epsilon}}{(2\epsilon +1)}\int_{0}^{\infty}\frac{\sigma^{2(\epsilon+1)}\mathrm{d}\sigma}{\left [ \sigma^{2}+(ma)^{2} \right ]^{\frac{1}{2}}\left [e^{2(\sigma^{2}+(ma)^{2})^{\frac{1}{2}}} + 1\right ]} \ .
\end{eqnarray}

The vacuum energy above provides the massless scalar field case by setting $m=0$. This gives
\begin{eqnarray}
\frac{\tilde{E_{C}}}{L^2} = -\frac{1}{4\pi^{2} a^{3}}\left ( \frac{l}{a} \right )^{2(\epsilon-1)}\frac{(-1)^{\epsilon}}{(2\epsilon +1)}\int_{0}^{\infty}\frac{\sigma^{2\epsilon+1}\mathrm{d}\sigma}{e^{2\sigma} + 1} \ .
\end{eqnarray}
Using the result of the integral below \cite{Grad}
\begin{eqnarray}
	\label{Int_1}
\int_{0}^{\infty}\frac{x^{\nu -1}\mathrm{d}x}{e^{\mu x} + 1}=\frac{1}{\mu ^{\nu }}\left ( 1-2^{1-\nu } \right )\Gamma (\nu )\zeta (\nu ) \ ,
\end{eqnarray}
where $\Gamma (\nu )$ corresponds to the Gamma function and $\zeta (\nu )$  is the Riemann zeta function, we arrive at
\begin{eqnarray}
\frac{\tilde{E_{C}}}{L^2} = -\frac{1}{4\pi^{2} a^{3}}\left ( \frac{l}{a} \right )^{2(\epsilon-1)}\frac{(-1)^{\epsilon}}{(2\epsilon +1)}\frac{\left ( 1-2^{-(2\epsilon +1)} \right )\Gamma (2\epsilon +2)\zeta (2\epsilon +2)}{2^{(2\epsilon +2)}} \ .
\end{eqnarray}
Consequently, we can exhibit two cases:
\begin{itemize}
\item{For $\epsilon=2$}:
\end{itemize}
\begin{eqnarray}
\frac{\tilde{E_{C}}}{L^2} = -\frac{31\pi^{4}}{322560 a^{3}}\left ( \frac{l}{a} \right )^{2} \ .
\end{eqnarray}

\begin{itemize}
\item{For $\epsilon=3$}:
\end{itemize}
\begin{eqnarray}
\frac{\tilde{E_{C}}}{L^2} = \frac{127\pi^{6}}{1720320 a^{3}}\left ( \frac{l}{a} \right )^{4} \ .
\end{eqnarray}

Because the integral in \eqref{energywithmassmpa} cannot be expressed in terms of elementary functions; let us evaluate its asymptotic limits. To do this, we make the following changes of variables $\xi^{2}=\sigma^{2}+(ma)^{2}$ and $\xi=mav$, so, we have
\begin{eqnarray}
\label{exactmpa}
\frac{\tilde{E_{C}}}{L^2} = -\frac{\left(am\right)^{2(\epsilon +1)}}{4\pi^{2} a^{3}}\left ( \frac{l}{a} \right )^{2(\epsilon-1)}\frac{(-1)^{\epsilon}}{(2\epsilon +1)}\int_{1}^{\infty}\frac{\left(v^{2}-1\right)^{\epsilon +\frac{1}{2}}\mathrm{d}v}{e^{2amv}+1} \ .
\end{eqnarray}
Knowing that the geometric series can be represented as
\begin{eqnarray}
	\label{geo_series_1}
\frac{1}{e^{2amv}+1}=\sum_{j=1}^{\infty}\left ( -1 \right )^{j+1}e^{-2amvj} \ ,
\end{eqnarray}
we obtain,
\begin{eqnarray}
\frac{\tilde{E_{C}}}{L^2} = -\frac{\left(am\right)^{2(\epsilon +1)}}{4\pi^{2} a^{3}}\left ( \frac{l}{a} \right )^{2(\epsilon-1)}\frac{(-1)^{\epsilon}}{(2\epsilon +1)}\sum_{j=1}^{\infty}\left ( -1 \right )^{j+1}\int_{1}^{\infty}\left(v^{2}-1\right)^{\epsilon +\frac{1}{2}}e^{-2amvj}\mathrm{d}v \ .
\end{eqnarray}
Finally, by using the integral representation of the modified Bessel function, the expression above becomes
\begin{eqnarray}
\frac{\tilde{E_{C}}}{L^2} = -\frac{\left(am\right)^{\epsilon +1}\Gamma \left ( \epsilon +\frac{3}{2} \right )(-1)^{\epsilon }}{4{\left (\pi \right )^{\frac{5}{2}}} a^{3}(2\epsilon +1)}\left ( \frac{l}{a} \right )^{2(\epsilon-1)}\sum_{j=1}^{\infty}\frac{\left ( -1 \right )^{j+1}}{j^{\epsilon +1} }K_{\epsilon +1}(2amj) \ .
\end{eqnarray}
We can now exam the asymptotic limits $ma>>1$ and $ma<<1$. They are:

$(i)$ For large arguments, $am>>1$, the modified Bessel can be expressed in an exponential form as shown in \eqref{Asympt_BesselK},  consequently the dominant term, $j=1$, provides
\begin{eqnarray}
\frac{\tilde{E_{C}}}{L^2} \approx  -\frac{\left(am\right)^{\epsilon+\frac{1}{2}}\Gamma \left ( \epsilon +\frac{3}{2} \right )(-1)^{\epsilon}}{8\pi^{2} a^{3}(2\epsilon +1)}\left ( \frac{l}{a} \right )^{2(\epsilon-1)}e^{-2am} \ .
\end{eqnarray}

$(ii)$ In order to analyze the case for $am<<1$ let us consider $\epsilon=2,3$:
\begin{itemize}
\item{In the case $\epsilon=2$, the expression \eqref{exactmpa} becomes}
\end{itemize}
\begin{eqnarray}
\frac{\tilde{E_{C}}}{L^2} = -\frac{\left(am\right)^{6}}{20\pi^{2} a^{3}}\left ( \frac{l}{a} \right )^{2}\int_{1}^{\infty}\frac{\left ( v^{2}-1\right )^{\frac{5}{2}}\mathrm{d}v}{e^{2amv}+1} \ .
\end{eqnarray}
By expanding the integrand in a series of positive powers of $v$, we can obtain an approximated expression for the LV Casimir energy per unity area, as shown below:
\begin{eqnarray}
\frac{\tilde{E_{C}}}{L^2} &\approx& -\frac{\left(am\right)^{6}}{20\pi^{2} a^{3}}\left ( \frac{l}{a} \right )^{2}\int_{1}^{\infty}\frac{\left ( v^{5}-\frac{5}{2}v^{3}+\frac{15}{8}v\right )\mathrm{d}v}{e^{2amv}+1}\nonumber\\&\approx&-\frac{31}{322560\pi^{2} a^{3}}\left ( \frac{l}{a} \right )^{2}\left [\pi^{6}-\frac{147}{31}\pi^{4}(am)^{2}+\frac{630}{31}\pi^{2}(am)^{4}  \right ] \ .
\end{eqnarray}
As to the pressure, we have,
\begin{eqnarray}
	{\tilde P}_C(a)=-\frac 1{322560a^4}\left ( \frac{l}{a} \right )^{2}\left[155\pi^4-441\pi^2(am)^2+630(am)^4\right].
\end{eqnarray}

\begin{itemize}
\item{In the case $\epsilon=3$, the expression \eqref{exactmpa} becomes}:
\end{itemize}
\begin{eqnarray}
\frac{\tilde{E_{C}}}{L^2} = \frac{\left(am\right)^{8}}{28\pi^{2} a^{3}}\left ( \frac{l}{a} \right )^{4}\int_{1}^{\infty}\frac{\left ( v^{2}-1\right )^{\frac{7}{2}}\mathrm{d}v}{e^{2amv}+1} \ .
\end{eqnarray}

\begin{figure}[h]
\subfigure[For $\epsilon=2$\label{Mpa2}]{\includegraphics[scale=0.35]{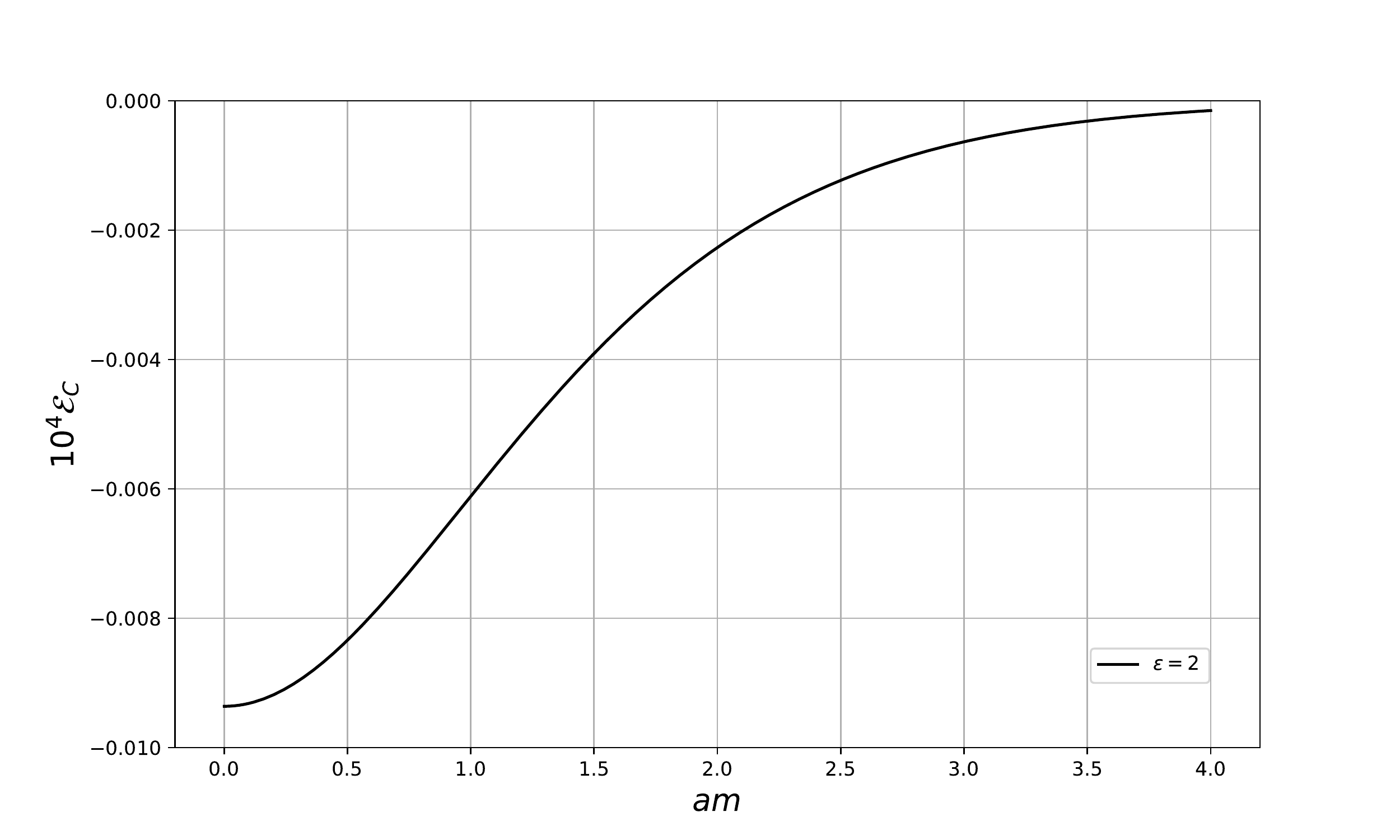}}
\subfigure[For $\epsilon=3$\label{Mpa3}]{\includegraphics[scale=0.35]{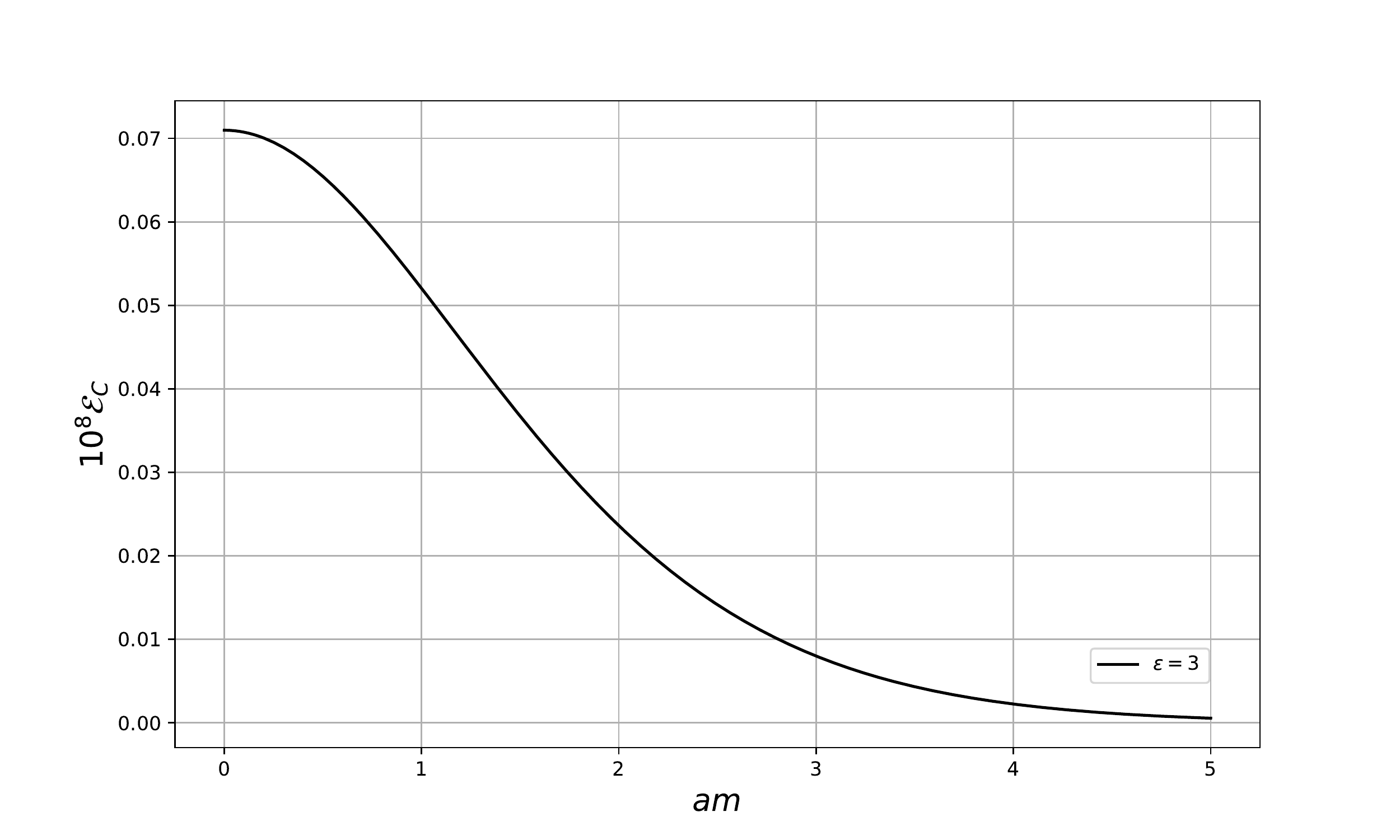}}
\caption{The Casimir energy in the case $u^{\mu}=(0,1,0,0)$ as a function of $am$, for mixed boundary condition. In this graph $\varepsilon _{c}=\frac{\tilde{E_{C}}}{L^2}a^{3}$, $\epsilon$ is equal to $2$ and $3$, in the left and right plots, respectively. For both plots we consider $\frac{l}{a}=0.01$.}
\label{fig4}
\end{figure}

Adopting a similar procedure as above, we have:
\begin{eqnarray}
\frac{\tilde{E_{C}}}{L^2} &\approx& \frac{\left(am\right)^{8}}{28\pi^{2} a^{3}}\left ( \frac{l}{a} \right )^{4}\int_{1}^{\infty}\frac{\left ( v^{7}-\frac{7}{2}v^{5}+\frac{35}{8}v^{3}-\frac{35}{16}v\right )\mathrm{d}v}{e^{2amv}+1}\nonumber\\&\approx&\frac{127}{1720320\pi^{2} a^{3}}\left ( \frac{l}{a} \right )^{4}\left [\pi^{8}-\frac{1240}{381}\pi^{6}(am)^{2}+\frac{980}{127}\pi^{4}(am)^{4}  \right ] \ .
\end{eqnarray}
In this case, the pressure is given by
\begin{eqnarray}
	{\tilde P}_C(a)=\frac 1{5160960a^4}\left ( \frac{l}{a} \right )^{4}\left[2667\pi^6-6200\pi^4(am)^2+8820\pi^2(am)^4\right],
\end{eqnarray}
In Fig \ref{fig4} we present the behavior of the Casimir energy per unit area as function of $ma$, considering {\bf again as an illustrative example} $\frac l a=0.01$, for two distinct values of $\epsilon$. We consider $\epsilon=2$ in the left plot, and $\epsilon=3$ in the right plot.

\subsubsection{Vector perpendicular to the plates}%%%%%%%%%%%%%%%%%%%%%%%%%%%%%%%%%%%%%%%%%%%%%%%%%%%%%%%%%%%%%%%%%%

Now, let us consider the four-vector $u^{\mu}$ as being perpendicular to the plates by choosing 
\begin{eqnarray}
u^{\mu}=(0,0,0,1) \ .
\end{eqnarray}
For  this case the corresponding dispersion relation is
\begin{eqnarray}
{\omega^2_{\textbf{k},n}} = k_{x}^2 + k_{y}^2 + \left [ (n+1/2)\frac{\pi}{a} \right ]^{2} + l^{2(\epsilon-1)}(-1)^{\epsilon}\left [ (n+1/2)\frac{\pi}{a} \right ]^{2\epsilon } + m^2 \ .
\end{eqnarray}
Consequently, the Hamiltonian operator, $\hat{H}$, is now given by
\begin{eqnarray}
\hat{H} = \frac{1}{2}\int\mathrm{d}^2\textbf{k} \sum_{n=0}^{\infty}{\omega_{\textbf{k},n}}\left[2 \hat{a}^{\dagger}_{\textbf{k,n}} \hat{a}_{\textbf{k,n}} + 
\frac{L^2}{(2\pi)^2}\right] \   .
\end{eqnarray}
The corresponding vacuum energy in this case is given by 
\begin{eqnarray}
E_{0} = \bra{0}\hat{H}\ket{0} = \frac{L^2}{8 \pi^2}\int \mathrm{d}^2\textbf{k} 
\sum_{n=0}^{\infty}{\omega_{\textbf{k},n}} \  .
\end{eqnarray}

Performing a change of coordinates in the plane $(k_{x},k_{y})$ to polar ones, a change of variable $u=ak$ and doing an expansion in $\frac{l}{a}<<1$, we obtain
\begin{eqnarray}
E_{0}&\approx&\frac{L^2}{8\pi^2a^{3}}\int_{0}^{2\pi}\mathrm{d}\theta\int_{0}^{\infty}u\mathrm{d}u\sum_{n=0}^{\infty}\Bigg\{\left[u^{2}+\left [ (n+1/2)\pi \right ]^{2}+(ma)^{2}\right ]^{\frac{1}{2}}\nonumber\\ &+&\frac{1}{2}\left(\frac{l}{a}\right)^{2(\epsilon-1)}(-1)^{\epsilon}\left [ \left ( n+1/2 \right )\pi \right ]^{2\epsilon }\left [ u^{2}+\left [ (n+1/2)\pi \right ]^{2}+(ma)^{2}\right ]^{-\frac{1}{2}}\Bigg\}\ .
\end{eqnarray}
The first term on the right hand side is associated with energy without Lorentz violation. Thus, after integration over the angular coordinate, the second term becomes
\begin{eqnarray}
\label{energywhioutviolationmper}
\tilde{E_{0}}=\frac{L^2(-1)^{\epsilon}}{8 \pi a^{3}}\left ( \frac{l}{a} \right )^{2(\epsilon-1)}\int_{0}^{\infty}u\mathrm{d}u
\sum_{n=0}^{\infty}\left [ (n+1/2)\pi \right ]^{2\epsilon }\left[u^{2}+\left [ (n+1/2)\pi \right ]^{2}+(ma)^{2}\right ]^{-\frac{1}{2}} \  .
\end{eqnarray}
By using the formula \eqref{Abel-half-integer} again, we find
\begin{eqnarray}
\label{energywhithabelmper}
\tilde{E_{0}} = \frac{L^2(-1)^{\epsilon }}{8 \pi a^{3}}\left ( \frac{l}{a} \right )^{2(\epsilon-1)}\int_{0}^{\infty}u\mathrm{d}u\Bigg\{\int_{0}^{\infty} 
F(t)\mathrm{d}t - i\int_{0}^{\infty}\frac{F(it)-F(-it)}{e^{2\pi t}+1}\mathrm{d}t\Bigg\}\   ,
\end{eqnarray}
where 
\begin{eqnarray}
F\left ( n+\frac{1}{2} \right ) = \left [ (n+1/2)\pi \right ]^{2\epsilon }\left[u^{2}+\left [ (n+1/2)\pi \right ]^{2}+(ma)^{2}\right ]^{-\frac{1}{2}} \  .
\end{eqnarray}
Note that the integral in the first term on the right hand side of Eq. \eqref{energywhithabelmper} is the free vacuum contribution, while the integral in the second term gives the renormalized vacuum energy per unit area, i.e.,
\begin{eqnarray}
\frac{\tilde{E_{C}}}{L^2} &=& -\frac{i}{8 \pi a^{3}}\left ( \frac{l}{a} \right )^{2(\epsilon-1)}\int_{0}^{\infty}u\mathrm{d}u\nonumber\\&\times&\int_{0}^{\infty}\mathrm{d}t \frac{(t\pi)^{2\epsilon }[u^2 + 
(it\pi)^{2}+(ma)^{2}]^{-1/2} -(t\pi)^{2\epsilon }[u^2 + 
(-it\pi)^{2}+(ma)^{2}]^{-1/2}}{e^{2 \pi t} + 1} \  .
\end{eqnarray}
By making the change of variable $t\pi = v$, we find
\begin{eqnarray}
\frac{\tilde{E_{C}}}{L^2}&=&-\frac{i}{8 \pi a^{3}}\left ( \frac{l}{a} \right )^{2(\epsilon-1)}\int_{0}^{\infty}u\mathrm{d}u\nonumber\\&\times&\int_{0}^{\infty}\mathrm{d}t \frac{v^{2\epsilon }\Bigg\{[u^2 + 
(iv)^{2}+(ma)^{2}]^{-1/2} - [u^2 + 
(-iv)^{2}+(ma)^{2}]^{-1/2}\Bigg\}}{e^{2v} + 1} \  . 
\end{eqnarray}

Again, analyzing the integral on the variable $v$ over the two intervals: $\left [ u^2 +(ma)^{2} \right ]^{1/2}>v$ and $\left [ u^2 +(ma)^{2} \right ]^{1/2}<v$, we obtain
\begin{eqnarray}
\frac{\tilde{E_{C}}}{L^2} = -\frac{1}{4\pi^{2} a^{3}}\left ( \frac{l}{a} \right )^{2(\epsilon-1)}\int_{0}^{\infty}u\mathrm{d}u\int_{\left [ u^{2}+(ma)^{2} \right ]^{\frac{1}{2}}}^{\infty}\mathrm{d}v\frac{v^{2\epsilon }[v^2-(u^{2}+(ma)^{2})]^{-1/2}}{e^{2v} + 1} \ .
\end{eqnarray}
By making and additional change of variable $\rho^{2} =v^{2}-(u^{2}+(ma)^{2})$, and transforming the coordinates in the plane  $(u,\rho)$ to polar ones, we arrive at
\begin{eqnarray}
\label{energywithmassmper}
\frac{\tilde{E_{C}}}{L^2} = -\frac{1}{4\pi^{2} a^{3}}\left ( \frac{l}{a} \right )^{2(\epsilon-1)}\int_{0}^{\infty}\frac{\left [ \sigma ^{2}+(ma)^{2} \right ]^{\epsilon -\frac{1}{2}}\sigma^{2}\mathrm{d}\sigma}{e^{2(\sigma^{2}+(ma)^{2})^{\frac{1}{2}}} + 1} \ .
\end{eqnarray}

For the case of the massless field, we take $m=0$. In this case, the integral reads,
\begin{eqnarray}
\frac{\tilde{E_{C}}}{L^2} = -\frac{1}{4\pi^{2} a^{3}}\left ( \frac{l}{a} \right )^{2(\epsilon-1)}\int_{0}^{\infty}\frac{\sigma^{2\epsilon+1}\mathrm{d}\sigma}{e^{2\sigma} + 1} \ .
\end{eqnarray}
By considering the same integral as in the last subsection, we find
\begin{eqnarray}
\frac{\tilde{E_{C}}}{L^2} = -\frac{1}{4\pi^{2} a^{3}}\left ( \frac{l}{a} \right )^{2(\epsilon-1)}\frac{\left ( 1-2^{-(2\epsilon +1)} \right )\Gamma (2\epsilon +2)\zeta (2\epsilon +2)}{2^{(2\epsilon +2)}} \ .
\end{eqnarray}
Let us analyze the cases for which $\epsilon=2,3$:
\begin{itemize}
\item{For $\epsilon=2$}:
\end{itemize}
\begin{eqnarray}
\frac{\tilde{E_{C}}}{L^2} = -\frac{31\pi^{4}}{64512 a^{3}}\left ( \frac{l}{a} \right )^{2} \ .
\end{eqnarray}

\begin{itemize}
\item{For $\epsilon=3$}:
\end{itemize}
\begin{eqnarray}
\frac{\tilde{E_{C}}}{L^2} = -\frac{127\pi^{6}}{245760 a^{3}}\left ( \frac{l}{a} \right )^{4} \ .
\end{eqnarray}

Once again, let us evaluate the integral in \eqref{energywithmassmper} in its asymptotic limits. To do this, we make the following changes of variables $\xi^{2}=\sigma^{2}+(ma)^{2}$ and $\xi=mav$, so, we obtain
\begin{eqnarray}
\label{exactmper}
\frac{\tilde{E_{C}}}{L^2} = -\frac{\left(am\right)^{2(\epsilon +1)}}{4\pi^{2} a^{3}}\left ( \frac{l}{a} \right )^{2(\epsilon-1)}\int_{1}^{\infty}\frac{v^{2\epsilon }\left(v^{2}-1\right)^{\frac{1}{2}}\mathrm{d}v}{e^{2amv}+1} \ .
\end{eqnarray}
Expressing the denominator in a geometric series as given in \eqref{geo_series_1}, we obtain
\begin{eqnarray}
\frac{\tilde{E_{C}}}{L^2} = -\frac{\left(am\right)^{2(\epsilon +1)}}{4\pi^{2} a^{3}}\left ( \frac{l}{a} \right )^{2(\epsilon-1)}\sum_{j=1}^{\infty}\left ( -1 \right )^{j+1}\int_{1}^{\infty}v^{2\epsilon }\left(v^{2}-1\right)^{\epsilon +\frac{1}{2}}e^{-2amvj}\mathrm{d}v \ .
\end{eqnarray}
In addition, using the identity below, 
\begin{eqnarray}
\frac{1}{\left ( 2mj \right )^{2\epsilon }}\frac{\mathrm{d^{2\epsilon }} \left ( e^{-2amvj} \right )}{\mathrm{d} a^{2\epsilon }}=v^{2\epsilon }e^{-2amvj} \ ,
\end{eqnarray}
the LV Casimir energy can be expressed as
\begin{eqnarray}
\frac{\tilde{E_{C}}}{L^2} = -\frac{\left(am\right)^{2(\epsilon +1)}}{4\pi^{2} a^{3}}\left ( \frac{l}{a} \right )^{2(\epsilon-1)}\frac{1}{\left ( 2m \right )^{2\epsilon }}\sum_{j=1}^{\infty}\frac{\left ( -1 \right )^{j+1}}{j^{2\epsilon }}\frac{\mathrm{d^{2\epsilon }} }{\mathrm{d} a^{2\epsilon }}\int_{1}^{\infty}\left(v^{2}-1\right)^{\frac{1}{2}}e^{-2amvj}\mathrm{d}v \ .
\end{eqnarray}

Using the integral representation for the modified Bessel function, the above expression becomes
\begin{eqnarray}
\frac{\tilde{E_{C}}}{L^2} = -\frac{\left(am\right)^{2(\epsilon +1)}}{4\pi^{2} a^{3}}\left ( \frac{l}{a} \right )^{2(\epsilon-1)}\frac{1}{(2m)^{2\epsilon +1}}\sum_{j=1}^{\infty}\frac{\left ( -1 \right )^{j+1}}{j^{2\epsilon +1}}\frac{\mathrm{d^{2\epsilon }} }{\mathrm{d} a^{2\epsilon }}\left ( \frac{K_{1}(2amj)}{a} \right ) \ .
\end{eqnarray}
Let us now analyze the asymptotic limits for $ma>>1$ and $ma<<1$:

$(i)$ For large arguments, $am>>1$, we can use the asymptotic form for the modified Bessel function, Eq. \eqref{Asympt_BesselK}, and taking the dominant term, $j=1$, we obtain that
\begin{eqnarray}
\frac{\tilde{E_{C}}}{L^2} \approx  -\frac{\left(am\right)^{2(\epsilon+1)}}{8\left ( \pi \right )^{\frac{3}{2}} a^{3}m^{\frac{1}{2}}(2m)^{2\epsilon +1}}\left ( \frac{l}{a} \right )^{2(\epsilon-1)}\frac{\mathrm{d^{2\epsilon }} }{\mathrm{d} a^{2\epsilon }}\left ( \frac{e^{-2am}}{a^{\frac{3}{2}}} \right ) \ .
\end{eqnarray}
We additionally want to consider the two cases $\epsilon=2,3$:
\begin{itemize}
\item{For case $\epsilon=2$}:
\end{itemize}
\begin{eqnarray}
\frac{\tilde{E_{C}}}{L^2} \approx -\frac{\left(am\right)^{\frac{9}{2}}}{16\left ( \pi \right )^{\frac{3}{2}}a^{3}}\left ( \frac{l}{a} \right )^{2}e^{-2am} \ .
\end{eqnarray}

\begin{itemize}
\item{For case $\epsilon=3$}:
\end{itemize}
\begin{eqnarray}
\frac{\tilde{E_{C}}}{L^2} \approx -\frac{\left(am\right)^{\frac{13}{2}}}{16\left ( \pi \right )^{\frac{3}{2}}a^{3}}\left ( \frac{l}{a} \right )^{4}e^{-2am} \ .
\end{eqnarray}

$(ii)$ For $am<<1$:
\begin{itemize}
\item{In the case $\epsilon=2$, the expression \eqref{exactmper} becomes}:
\end{itemize}
\begin{eqnarray}
\frac{\tilde{E_{C}}}{L^2} = -\frac{\left(am\right)^{6}}{4\pi^{2} a^{3}}\left ( \frac{l}{a} \right )^{2}\int_{1}^{\infty}\frac{v^{4}\left ( v^{2}-1\right )^{\frac{1}{2}}\mathrm{d}v}{e^{2amv}+1} \ .
\end{eqnarray}

Expanding the integrand in powers of $v$, we obtain an expression that allows us to evaluate the integral:
\begin{eqnarray}
\frac{\tilde{E_{C}}}{L^2} &\approx& -\frac{\left(am\right)^{6}}{4\pi^{2} a^{3}}\left ( \frac{l}{a} \right )^{2}\int_{1}^{\infty}\frac{\left ( v^{5}-\frac{1}{2}v^{3}-\frac{1}{8}v\right )\mathrm{d}v}{e^{2amv}+1}\nonumber\\&\approx&-\frac{31}{64512\pi^{2} a^{3}}\left ( \frac{l}{a} \right )^{2}\left [\pi^{6}-\frac{147}{155}\pi^{4}(am)^{2}-\frac{42}{31}\pi^{2}(am)^{4}  \right ] \ .
\end{eqnarray}
As to the pressure, we have,
\begin{eqnarray}
	{\tilde P}_C(a)=-\frac 1{322560a^4}\left ( \frac{l}{a} \right )^{2}\left[775\pi^4-441\pi^2(am)^2-210(am)^4\right].
\end{eqnarray}

\begin{figure}[h]
\subfigure[For $\epsilon=2$\label{Mper2}]{\includegraphics[scale=0.35]{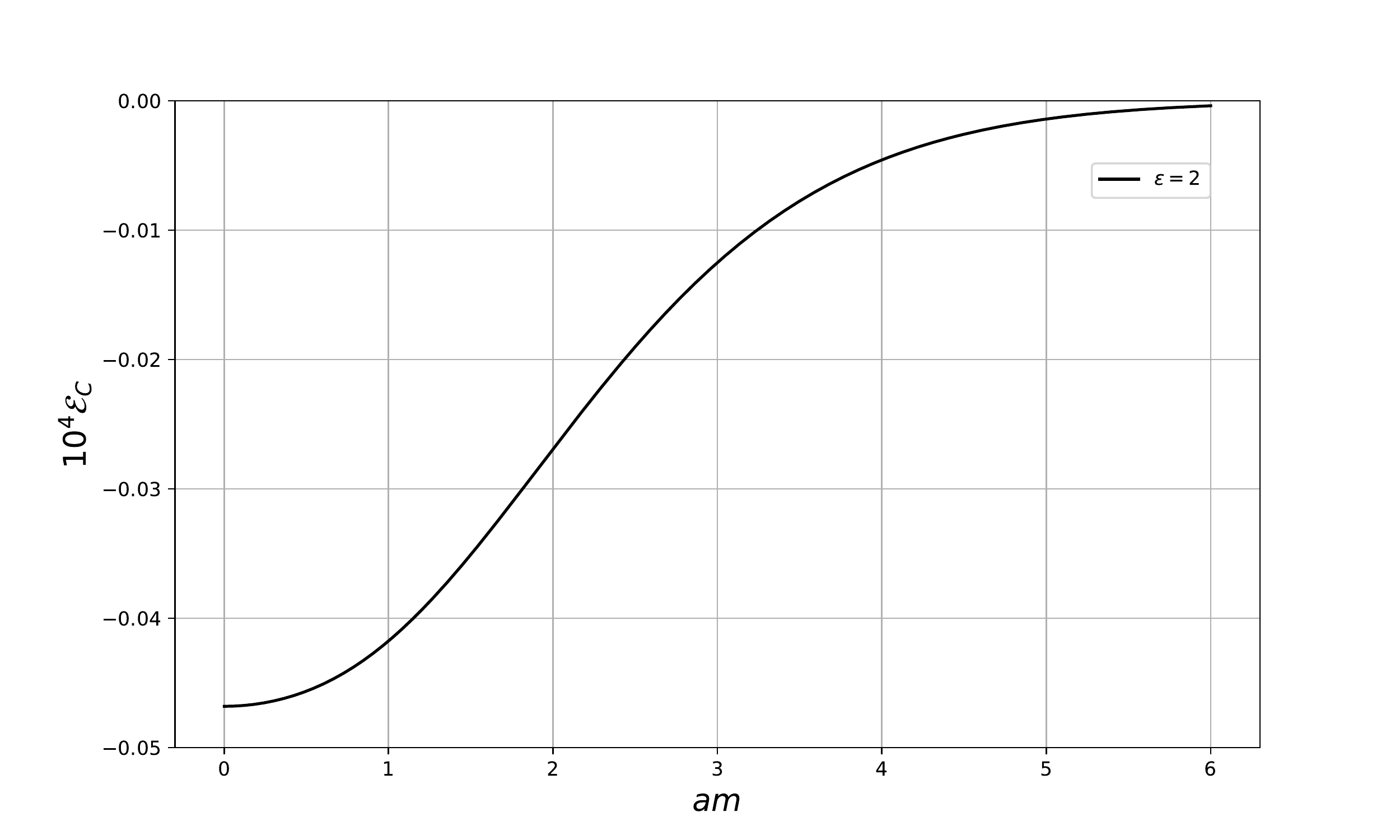}}
\subfigure[For $\epsilon=3$\label{Mper3}]{\includegraphics[scale=0.35]{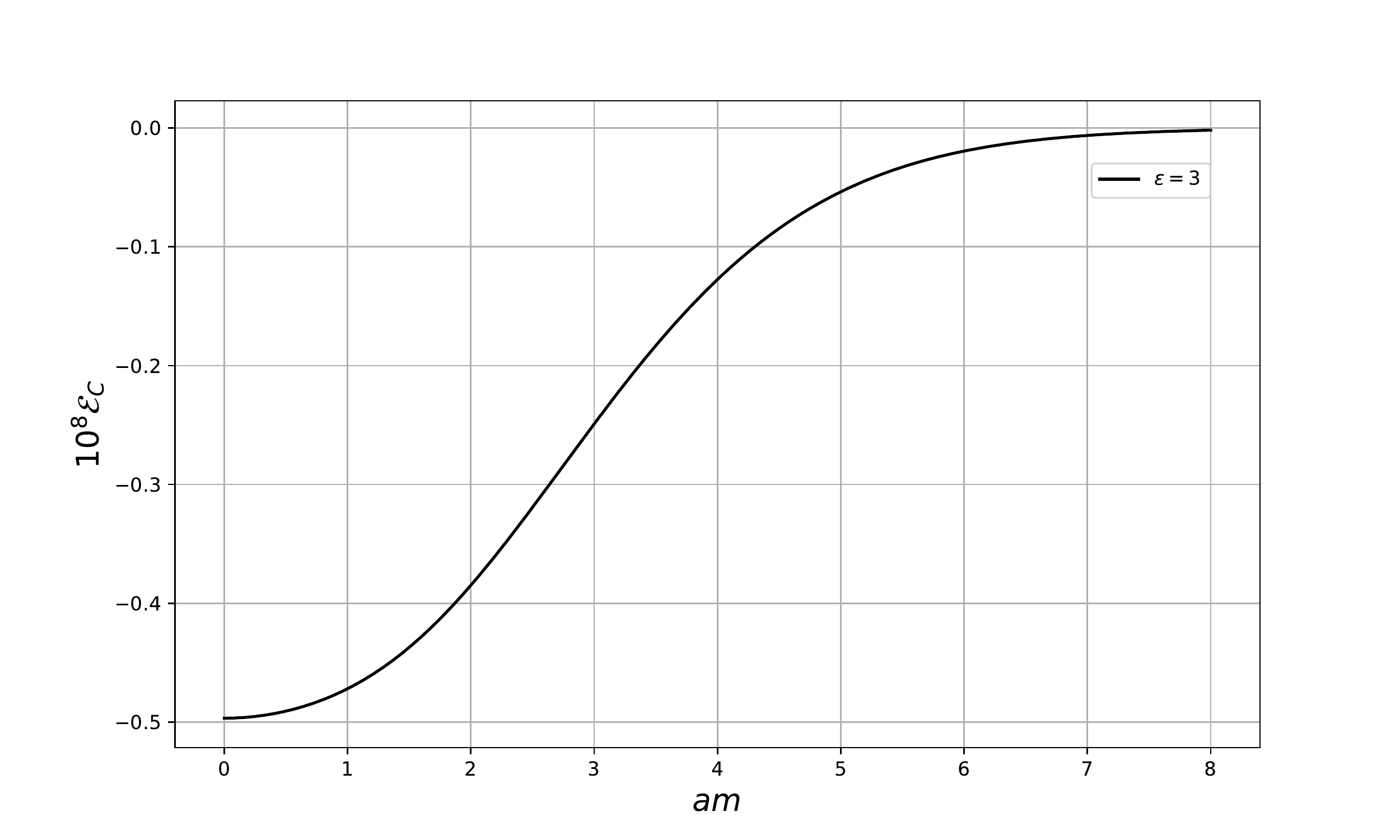}}
\caption{The Casimir energy in the case $u^{\mu}=(0,0,0,1)$ as a function of $am$, considering the mixed boundary condition obeyed by the fields. In this graph $\varepsilon _{c}=\frac{\tilde{E_{C}}}{L^2}a^{3}$ and $\frac{l}{a}=0.01$.}
\label{fig5}
\end{figure}

\begin{itemize}
	\item{In the case $\epsilon=3$, the expression \eqref{exactmper} becomes}:
\end{itemize}
\begin{eqnarray}
	\frac{\tilde{E_{C}}}{L^2} = -\frac{\left(am\right)^{8}}{4\pi^{2} a^{3}}\left ( \frac{l}{a} \right )^{4}\int_{1}^{\infty}\frac{v^{6}\left ( v^{2}-1\right )^{\frac{1}{2}}\mathrm{d}v}{e^{2amv}+1} \ ,
\end{eqnarray}
the series expansion in this case provides
\begin{eqnarray}
	\frac{\tilde{E_{C}}}{L^2} &\approx& -\frac{\left(am\right)^{8}}{4\pi^{2} a^{3}}\left ( \frac{l}{a} \right )^{4}\int_{1}^{\infty}\frac{\left ( v^{7}-\frac{1}{2}v^{5}-\frac{1}{8}v^{3}-\frac{1}{16}v\right )\mathrm{d}v}{e^{2amv}+1}\nonumber\\&\approx&-\frac{127}{245760\pi^{2} a^{3}}\left ( \frac{l}{a} \right )^{4}\left [\pi^{8}-\frac{1240}{2667}\pi^{6}(am)^{2}-\frac{28}{127}\pi^{4}(am)^{4}  \right ] \ .
\end{eqnarray}
Consequently, the pressure reads,
\begin{eqnarray}
	{\tilde P}_C(a)=-\frac 1{5160960a^4}\left ( \frac{l}{a} \right )^{4}\left[18669\pi^6-6200\pi^4(am)^2-1764\pi^2(am)^4\right]
\end{eqnarray}
{\bf Once more, an illustrative example in} Fig \ref{fig5} we present the behavior of the Casimir energy per unit of area as function of $ma$, considering $\frac l a=0.01$, for two distinct values of $\epsilon$. In the left panel, we consider $\epsilon=2$ while in the right panel $\epsilon=3$.

\section{Concluding Remarks}
\label{Concl}
In this work we have investigated the influence of the Lorentz symmetry violation on the Casimir energy associated to a real massive scalar quantum field. We considered the situation in which the field is confined between two parallel plates and assumed  that the field obeys boundary conditions of the types Dirichlet, Neumann and mixed on the plates of area $L^2$ separated by the distance $a\text{ }(a\ll L)$.

The Lorentz symmetry violation is implemented admitting a direct coupling between a constant space-like four vector, $u^\mu$, in a scenario aether-like CPT-even theoretical model,  with higher-order derivative of the  field, represented by $l^{2(\epsilon-1)}(u\partial)^{2\epsilon} \phi(x)$, as exhibited in the modified Klein-Gordon equation, Eq. \eqref{kgmodific}, being $\epsilon$ an integer number greater or equal to $2$, and $l$ is a parameter of order inverse the energy scale where the Lorentz symmetry is broken. As to the constant four-vector, two distinct directions are considered: the vector parallel to the plates and the vector perpendicular to them. We have verified that the combined modifications in the dynamic of the quantum field, produce important corrections on the corresponding dispersion relations, and consequently on the Casimir energies. In the calculation of the Casimir energy, we have to develop an integral over the bi-dimensional space in the $\vec{k}$ plane associated with the momentum of the field parallel to the plates, and a sum over discrete momentum orthogonal to them, as exhibited in Eq. \eqref{vacuumdpa}, for example. To develop the summation over the discrete momentum, we adopted the Abel-Plana summation formula for integer, Eq. \eqref{Abel}, and half-integer, Eq. \eqref{Abel-half-integer}, quantum numbers, respectively. Because the integrals involved in the obtainment of the Casimir energy do not provide very enlightening results, in our analysis we decided to develop an expansion on the parameter $\frac la<<1$ in the integrand, keeping up to the first order term. Adopting this procedure, we were able to provide the corrections on the Casimir energy and pressure due to the LV term, caused by specific boundary condition obeyed by the quantum field on the plates, the direction of the space-like constant vector and the order of the derivative. Because the dispersion relation for the Neumann boundary condition is analogous to the one in the Dirichlet condition, we only provide a brief discussion for this case.  By our results for the Casimir pressures induced by the LV term, $\tilde{P}_c(a)$, we would like to emphasize that negative values, that correspond to an attracted forces between the plates, and positive values, that correspond to a repulsive forces between the plates, depend on the direction of the constant vector, $u^\mu$, with respect to the plates, on the value assumed for the parameter $\epsilon$  and also depend on the boundary condition imposed on the field at the plates.  

In all our analysis, the LV Casimir energy is expressed in term of an integral representation for massive field. So, in order to  furnish some quantitative information about this energy, we provided its asymptotic expressions for $am>>1$ and $am<<1$. In the former, the Casimir energy decay exponentially as $e^{-2am}$,  while in the opposite limit it presents a term that corresponds to the massless case with additional corrections proportional to some power of the product $am$. In addition, we also have presented four graphs for the Casimir energies as function of $am$ considering $\epsilon$ equal to $2$ and $3$, contemplating all the possibles scenarios. Of course, the intensity of the LV Casimir energy depends on the order of the higher-derivative term. It is smaller for higher value of $\epsilon$.   Considering Dirichlet condition, and the vector parallel to the plates, the LV Casimir energy and pressure are positive for $\epsilon=2$ and negative for $\epsilon=3$; however for vector perpendicular to the plates, the LV Casimir energy and pressure present the same positive sign for both values of $\epsilon$. For mixed boundary condition, the same behavior related to the sign of the LV Casimir energies and pressure are observable. We would like to say that  there are changes in the sign for the Casimir energies when the field obeys the Dirichlet boundary condition and mixed one, for each specific situation. 

To finish this section we want to make a few comments about the results obtained in sections \ref{Dirichlet} and \ref{Mixed}. The corrections induced by the Lorentz violation in the Casimir pressure never vanish. Accepting that the Lorentz violation is part of the source in the $1\%$ experimental error estimated in \cite{Harris}, it is possible to infer an upper bound for the parameter $l$. Considering the distance $a$ between the parallel plates being of order $10^{-8}m$,  the upper limit for $l$ is of order $10^{-9}m$ for the case of $\epsilon=2$ and of order  $10^{-8}m$ for the case of $\epsilon=3$. Here in this paper we have analyzed the Casimir effect associated with scalar field in a general scenario of LV.  In this sense, the results obtained to the LV Casimir energies include two different ingredients: the presence of a background constant vector and spatial higher-order derivative terms. The analysis of the Casimir energy associated with electromagnetic fields in a LV scenario, has been developed in \cite{Deivid}. In this work it is discussed the corrections on the standard Casimir energy due to two different LV approaches.

\section*{Acknowledgment} 

We would like to thank A. Yu. Petrov for valuable discussions during the development of this work.  R.A.D  thanks Conselho Nacional de Desenvolvimento Científico e Tecnológico (CNPq.). H.F.S.M. is partially supported by CNPq. under the Grants no 311031/2020-0.  E.R.B.M is partially supported by CNPq under Grant no 301.783/2019-3.

\end{document}